\documentclass[
    reprint,         % Two-column format (like PRL)
    aps,             % American Physical Society style
    prl,             % Journal: Physical Review Letters
    superscriptaddress, % Author affiliations in superscripts
    showpacs,        % Show PACS numbers (optional)
    floatfix,        % Fixes for floating figures/tables
    nofootinbib      % Keeps footnotes out of bibliography
]{revtex4-2}

\usepackage[utf8]{inputenc}    % UTF-8 encoding
\usepackage[T1]{fontenc}       % Better font encoding
\usepackage{amsmath, amssymb}  % Math symbols
\usepackage{bm}                % Bold math
\usepackage{graphicx}          % Include graphics
\usepackage{xcolor}            % Color text
\usepackage{hyperref}          % Hyperlinks
\usepackage{physics}           % Common physics notations (\bra, \ket, etc.)
\usepackage{siunitx}           % SI units formatting
\usepackage{microtype}         % Better text spacing
\usepackage{newtxtext}         % PRL-like font
\usepackage{newtxmath}         % PRL-like font
\usepackage{multirow}          % For tables
\usepackage{threeparttable}    % For tables
\usepackage{soul}

\hypersetup{
    colorlinks=true,
    linkcolor=blue,
    citecolor=blue,
    urlcolor=blue,
    pdfauthor={Your Name},
    pdftitle={Your Title},
}

\begin{document}
%\title{Magneto-Optical Trapping of CaH Molecules}
\title{Magneto-Optical Trapping of a Metal Hydride Molecule}

\author{Jinyu Dai}
\email{jd3706@columbia.edu}
\affiliation{Department of Physics, Columbia University, New York, New York 10027, USA}

\author{Benjamin Riley}
\affiliation{Department of Physics, Columbia University, New York, New York 10027, USA}

\author{Qi Sun}
\affiliation{Department of Physics, Columbia University, New York, New York 10027, USA}

\author{Debayan Mitra}
\affiliation{Department of Physics, Columbia University, New York, New York 10027, USA}
\affiliation{Department of Physics, Indiana University, Bloomington, Indiana 47405, USA}

\author{Tanya Zelevinsky}
\affiliation{Department of Physics, Columbia University, New York, New York 10027, USA}

\date{\today}

\begin{abstract}
% \noindent
We demonstrate a three-dimensional magneto-optical trap (MOT) of a metal hydride molecule, CaH. We are able to scatter $\sim$$10^{4}$ photons with vibrational loss covered up to vibrational quantum number $\nu=2$. This allows us to laser slow the molecular beam near zero velocity with a ``white-light'' technique and subsequently load it into a radio-frequency MOT. The MOT contains $230(40)$ molecules, limited by beam source characteristics and predissociative loss of CaH. The temperature of the MOT is below one millikelvin. The predissociative loss mechanism could, in turn, facilitate controlled dissociation of the molecule, offering a possible route to optical trapping of hydrogen atoms for precision spectroscopy.
\end{abstract}

\maketitle

%%%%%%%%%%%%%%%%%%%%%%%%%%%%%%% Introduction %%%%%%%%%%%%%%%%%%%%%%%%%%%%%%%

%\section{Introduction}

The development of techniques to manipulate molecules in the ultracold regime has enabled a variety of applications in quantum computation~\cite{holland2023cafentangle,bao2023cafentangle,picard_entanglement_2025,ruttley_long-lived_2025}, ultracold chemistry~\cite{liu2022bimolecular,ZelevinskyMcDonaldNature16_Sr2PD,liu_hyperfine--rotational_2025}, and precision measurement~\cite{leung2023molecularclock,anderegg2023caohedm}. One notable technique is direct laser cooling of molecules, which, up to now, has led to several types of diatomic and polyatomic molecules being trapped in magneto-optical traps (MOTs)~\cite{barry2014magneto,truppe2017molecules,anderegg2017cafmot,Williams_2017,collopy20183d,vilas2022magneto,zeng2024three,lasner2025magneto,alfmot2025}. High phase-space densities can then be achieved with further cooling~\cite{burau2023bluemot,li_conveyor-belt_2025}. Extending these techniques to additional molecular species creates new opportunities. Metal hydride molecules are of particular interest, as they constitute a simple yet rich class of diatomic molecules for precise studies of molecular structure as well as ultracold dynamics and chemistry. Notably, cold and ultracold metal hydride molecules are of astrophysical significance, as they have been observed in stellar and interstellar media~\cite{Frum_rotational_spectra_1993,Alavi2018einsteincoeff}. Studying their chemical reactivity in a controlled laboratory environment would improve our understanding of fundamental chemical processes~\cite{tscherbul2020magnetic}. At the same time, they may offer a promising pathway toward producing an ultracold and trapped sample of atomic hydrogen~\cite{lane2015hydrogen,vazquez2022direct,sun2023probing}, an ideal system for high-precision tests of quantum electrodynamics and measurements of fundamental constants~\cite{tiesinga2021codata,biraben_spectroscopy_2009}.

Hydrogen spectroscopy has advanced remarkably over the past few decades. The narrow $1S$--$2S$ transition was measured with kHz- to Hz-level precision~\cite{cesar1996two,parthey2011improved}, and other transitions, including $2S$--$4P$~\cite{beyer2017rydberg}, $2S_{1/2}$--$2P_{1/2}$ (Lamb shift)~\cite{bezginov2019lamb}, $1S$--$3S$~\cite{grinin2020two}, and $2S_{1/2}$--$8D_{5/2}$~\cite{brandt2022measurement} transitions, were also measured with high precision to help refine the proton charge radius~\cite{tiesinga2021codata}. These measurements are limited by velocity-dependent broadening arising from the finite temperature of the atomic beam. Further improvements could be achieved by producing a trapped, ultracold sample of hydrogen atoms. While laser cooling of hydrogen remains technically challenging, and a magnetic trap environment is unsuitable for high-precision measurements~\cite{fried1998bose}, an approach based on dissociation of ultracold metal hydride molecules could be feasible \cite{sun2023probing}.

Diatomic alkaline-earth metal hydride molecules possess an internal structure that allows for efficient laser cooling~\cite{di_rosa_laser-cooling_2004,mcnally_optical_2020,vazquez2022direct}. Following near-threshold dissociation, the resulting hydrogen atoms could reach a lower temperature than the parent molecules~\cite{lane2015hydrogen,sun2023probing}. The initial challenge of this approach is to trap an ultracold molecular sample. In this Letter, we show that calcium monohydride (CaH) can be laser slowed and loaded into a MOT. We trap $\sim$$200$~molecules in the MOT, at submillikelvin temperatures. The MOT characteristics, including the lifetime and effective trap frequency, are measured and found to be similar to those of other molecular MOTs. The molecule number is primarily limited by the beam source characteristics, with a relatively high forward velocity and an order-of-magnitude lower yield than for diatomic metal fluorides~\cite{sun2025chemistry}. The predissociative loss of CaH further reduces the number of trappable molecules~\cite{sun2023probing}. This achievement marks an important milestone toward a new platform for studies of quantum chemistry in the ultracold regime and optical trapping of atomic hydrogen.

%%%%%%%%%%%%%%%%%%%%%%%%%%%%%%% CaH laser slowing %%%%%%%%%%%%%%%%%%%%%%%%%%%%%%%

% \section{C\MakeLowercase{a}H laser slowing}

\begin{figure}[t]
   \centering
   \includegraphics[scale=0.4]{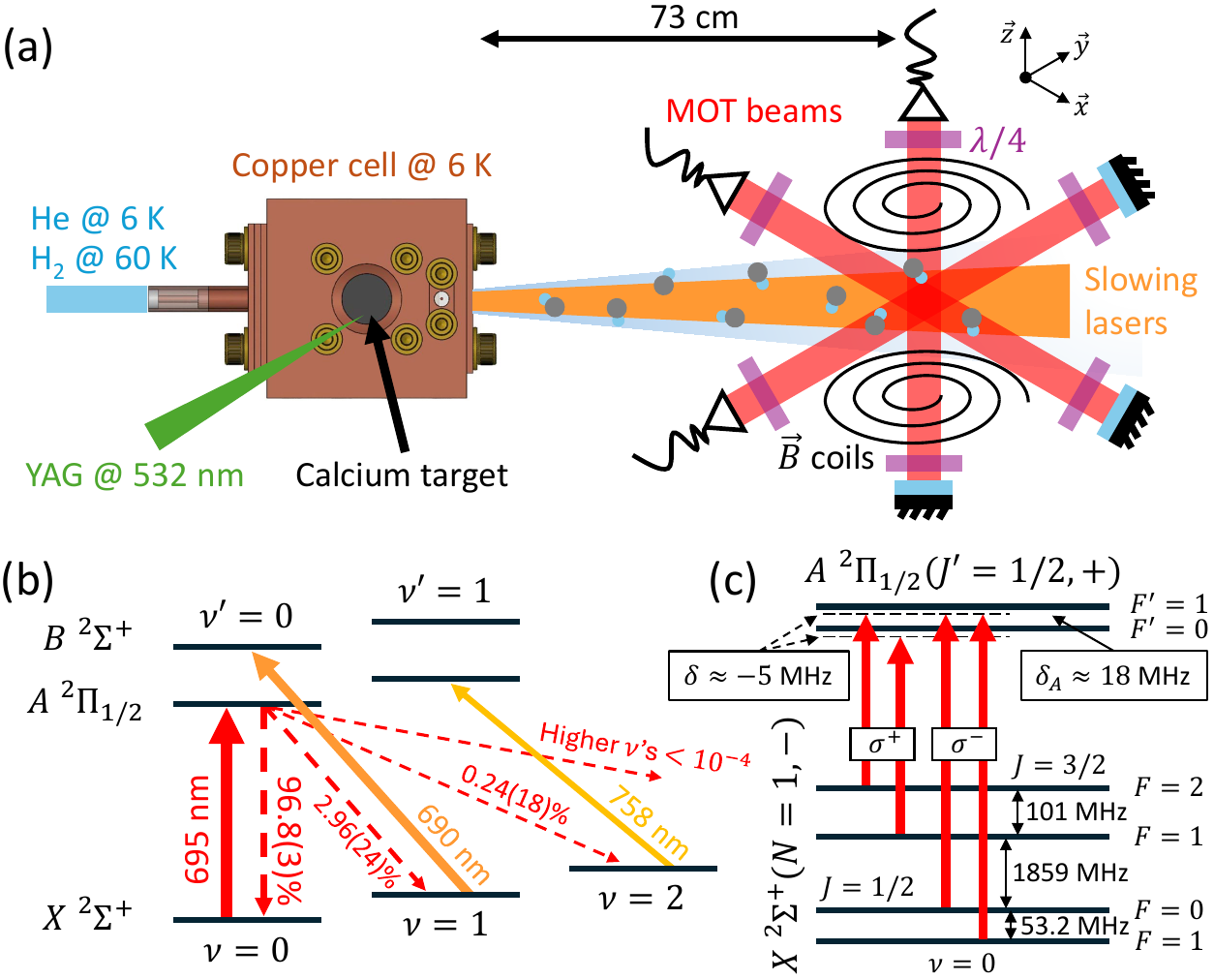}
   \caption{(a) Schematic of the experiment. CaH molecules are generated from a CBGB source operating at $\sim$$6$~K. The molecules are subsequently laser slowed and trapped in the MOT region $73$~cm away from the cell exit. (b) Relevant energy level structure for laser slowing. The deceleration force is supplied by the frequency-broadened main cycling laser addressing the $A^{2}\Pi_{1/2}(\nu'=0,\ J'=1/2,\ +)\leftarrow X^{2}\Sigma^{+}(\nu=0,\ N=1,\ -)$ transition at $695$~nm. Vibrational repumping covers up to ($\nu=2$) for both laser slowing and the MOT. The shown VBRs were experimentally measured in Ref.~\cite{vazquez2022direct}. (c) Relevant energy level structure for the MOT. The lasers address the same transition as the main cycling laser of slowing but with a distinct laser frequency for each hyperfine state. The $A^2\Pi_{1/2}(\nu'=0,\ J'=1/2,\ +)$ excited state has a $\delta_A\approx18$~MHz hyperfine splitting. The MOT is optimal with a global detuning of $\delta\approx-5$~MHz.}
   \label{fig:schematic}
\end{figure}

The experiment starts with a cryogenic buffer-gas beam (CBGB)~\cite{barry_bright_2011,hutzler_buffer_2012,bulleid_characterization_2013} of CaH molecules operating at $\sim$$6$~K, as shown in Fig.~\hyperref[fig:schematic]{\ref*{fig:schematic}(a)}. A pulsed Nd:YAG laser at $532$~nm ablates a solid calcium target, creating a hot plume of Ca atoms. The atoms then react with hydrogen molecules (H$_{2}$) introduced through a thermally isolated fill line at $\sim$$60$~K. The product CaH molecules are further buffer-gas cooled with $^{4}$He at $\sim$$6$~K and extracted from the cell to form the molecular beam. The forward velocity peaks at $\sim$$300$~m$/$s. With its high vapor pressure at cryogenic temperatures, H$_{2}$ can act as a buffer gas, helping the production of cold molecules at the cost of an increased forward velocity due to its light mass~\cite{sun2025chemistry}. Buffer-gas cooling with $^{4}$He extends the low-velocity tail of the distribution to $\sim$$100$~m$/$s, providing the basis for efficient laser slowing and trapping.

Counter-propagating slowing lasers are applied to the molecular beam extracted from the cell. The slowing lasers include a main cycling laser addressing the $A^{2}\Pi_{1/2}(\nu'=0,\ J'=1/2,\ +)\leftarrow X^{2}\Sigma^{+}(\nu=0,\ N=1,\ -)$ transition and two vibrational repumping lasers for ($\nu=1$) and ($\nu=2$) addressing the $B^{2}\Sigma^{+}(\nu'=0,\ N'=0,\ +)\leftarrow X^{2}\Sigma^{+}(\nu=1,\ N=1,\ -)$ and the $A^{2}\Pi_{1/2}(\nu'=1,\ J'=1/2,\ +)\leftarrow X^{2}\Sigma^{+}(\nu=2,\ N=1,\ -)$ transitions. Here, $\nu$, $N$, $J$, and $+/-$ denote the vibrational quantum number, total angular momentum excluding spin, total angular momentum excluding nuclear spin, and state parity, respectively. The relevant energy level structure is shown in Fig.~\hyperref[fig:schematic]{\ref*{fig:schematic}(b)}. The deceleration force is provided by the main cycling laser at $695$~nm with its $96.8$\% vibrational branching ratio (VBR) back to the ($\nu=0$) state. To decelerate the slowest molecules with a forward velocity of $\sim$$100$~m$/$s to the MOT capture velocity ($\lesssim$$10$~m$/$s), at least $\sim$$7\times 10^{3}$~photons must be scattered, given the $\hbar k/m\approx 1.4\times 10^{-2}$~m$/$s recoil velocity per photon. Here, $\hbar$ is the reduced Planck constant, $k$ is the wave number of a $695$~nm photon, and $m$ is the mass of a CaH molecule. Based on the theoretical values in Ref.~\cite{dai2024laser}, adding vibrational repumps up to ($\nu=2$) allows for $1/(1-\sum_{\nu=0}^2\text{VBR}_{0\rightarrow\nu})\sim5\times10^{4}$~photons to be scattered before $<$$37$\% ($<$$1/e$) of the initial population remains. Although VBRs for the higher vibrational states have not been measured, and theoretical values could be underestimated~\cite{vazquez2022direct,sun2023probing}, this order-of-magnitude higher photon budget should be adequate for observing a MOT.

We employ white-light laser slowing~\cite{hemmerling_laser_2016}, where all the slowing lasers are spectrally broadened with strongly driven electro-optic modulators (EOMs) operating at $\sim$$4.3$~MHz. Additional EOMs are used to cover the $\sim$$1.9$~GHz spin-rotation splitting and the $\sim$$50$--$100$~MHz hyperfine splittings [Fig.~\hyperref[fig:schematic]{\ref*{fig:schematic}(c)}]. The strongly driven white-light EOM, together with the hyperfine EOM, produce a frequency broadening of $\sim$$400$~MHz. Further details of the laser configuration can be found in Supplemental Material~\cite{Supplemental_Material}. Molecules with velocities up to $\sim$$200$~m$/$s thus remain resonant with the slowing lasers and are decelerated continuously. To avoid populating magnetic dark states during the slowing process, we use a Pockels cell to switch the slowing lasers between two orthogonal linear polarizations at a rate of $2$~MHz. The slowing lasers are gently focused toward the cell to compensate for the divergence of the molecular beam, with $0.8$~cm $1/e^2$ Gaussian diameter at the cell exit and $1.3$~cm $1/e^2$ Gaussian diameter at the MOT center which is $73$~cm downstream from the cell [Fig.~\hyperref[fig:schematic]{\ref*{fig:schematic}(a)}]. The total powers used for the slowing lasers are $1.2$, $0.9$, and $0.3$~W, corresponding to the main cycling laser, ($\nu=1$) repumping laser, and ($\nu=2$) repumping laser, respectively.

\begin{figure}[t]
   \centering
   \includegraphics[scale=0.41]{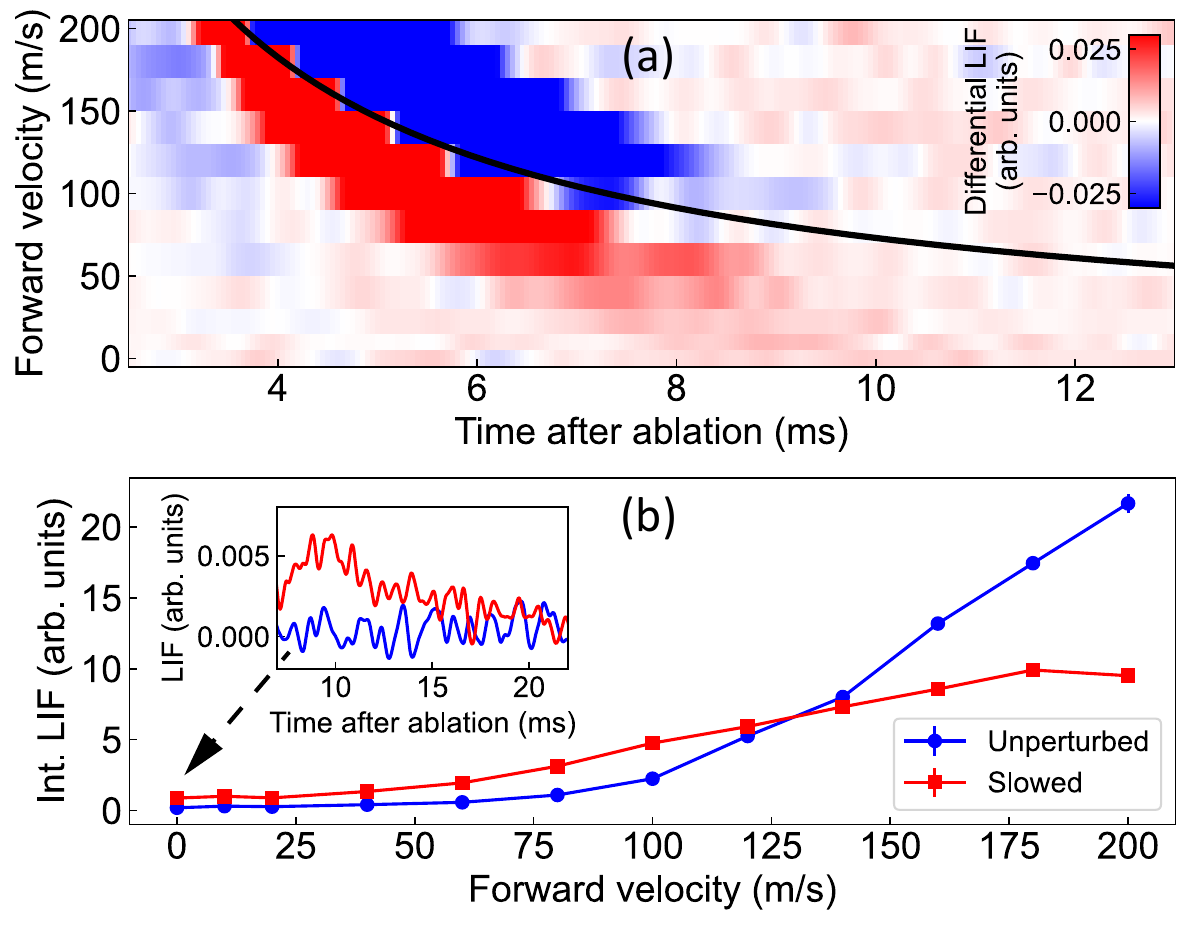}    
   \caption{Characterization of laser slowing. (a) LIF of an unperturbed molecular beam subtracted from the LIF of a molecular beam subject to laser slowing. The solid black line marks ballistic propagation from the cell and serves as a guide to the eye. (b) Time-integrated LIF for the unperturbed and slowed beams as a function of forward velocity. Error bars represent 1-$\sigma$ uncertainties and are too small to be visible on this scale. The inset shows fluorescence for detection at zero velocity. The results indicate efficient laser slowing, with molecules decelerated below the MOT capture velocity ($\lesssim$$10$~m$/$s).}
   \label{fig:Slowing}
\end{figure}

We characterize the effectiveness of laser slowing via the change in forward velocity measured using a two-photon background-free detection scheme~\cite{barry_slowing_2012}. We apply a laser transversely to the molecular beam in the MOT region to address the $A^2\Pi_{1/2}(\nu'=0,\ J'=3/2,\ +)\leftarrow X^2\Sigma^+(\nu=0,\ N=1,\ -)$ and the $A^2\Pi_{1/2}(\nu'=0,\ J'=3/2,\ +)\leftarrow X^2\Sigma^+(\nu=0,\ N=3,\ J=5/2,\ -)$ transitions at $695$~nm and apply a second laser in a velocity-sensitive configuration---copropagating with the slowing lasers---to address the $E^2\Pi_{1/2}(\nu'=0,\ J'=3/2,\ -)\leftarrow A^2\Pi_{1/2}(\nu=0,\ J=3/2,\ +)$ transition at $1668$~nm. The $490$~nm laser-induced fluorescence (LIF) from the $E\rightarrow X$ decay is collected on a photomultiplier tube (PMT) with temporal resolution, and the velocity information is obtained through detuning the $E\leftarrow A$ laser to compensate for the corresponding Doppler shift. In this approach, the $A$ state chosen for detection is not involved in the slowing scheme. Laser slowing is found to be optimal when it is turned on $1.2$~ms after ablation (when the molecules are fully extracted from the cell) and with a duration of $15$~ms. Figure~\ref{fig:Slowing} shows the slowing measurements in this configuration. The differential LIF in Fig.~\hyperref[fig:Slowing]{\ref*{fig:Slowing}(a)} is obtained by subtracting the fluorescence of an unperturbed molecular beam from that of a molecular beam subject to laser slowing. The time-integrated LIF signals versus molecular velocity are shown in Fig.~\hyperref[fig:Slowing]{\ref*{fig:Slowing}(b)}. Given the signal-to-noise ratio (SNR), each measurement shown in this Letter requires averaging $\sim$$500$ experimental cycles. The results show the overall distribution shifting to lower velocities with laser slowing. In particular, with slowing we observe molecules near zero velocity ($E\leftarrow A$ laser on resonance), as shown in the inset of Fig.~\hyperref[fig:Slowing]{\ref*{fig:Slowing}(b)}. This indicates the presence of molecules within the MOT capture velocity.

%%%%%%%%%%%%%%%%%%%%%%%%%%%%%%% CaH MOT %%%%%%%%%%%%%%%%%%%%%%%%%%%%%%%

% \section{C\MakeLowercase{a}H MOT}

The MOT lasers address the same main cycling transition. However, since the MOT can work only within the natural linewidth [$\Gamma/(2\pi)\sim 5$~MHz], each hyperfine state is addressed with a distinct laser frequency [Fig.~\hyperref[fig:schematic]{\ref*{fig:schematic}(c)}]. CaH has an $18$~MHz hyperfine splitting in the $A^2\Pi_{1/2}(\nu'=0,\ J'=1/2,\ +)$ excited state, which is comparable to the hyperfine splitting reported in the other parity state \cite{di_rosa_laser-cooling_2004}. Given the hyperfine branching ratios, $X^2\Sigma^+(\nu=0,\ N=1,\ J=3/2,\ F=1,\ -)$ is coupled to the $A^2\Pi_{1/2}(\nu'=0,\ J'=1/2,\ F'=0,\ +)$ state, while the other three hyperfine levels in the ground state are coupled to $A^2\Pi_{1/2}(\nu'=0,\ J'=1/2,\ F'=1,\ +)$. All the MOT laser beams have $1/e^2$ Gaussian diameters of $1.7$~cm and roughly equal intensities. The lasers are globally detuned by $-5$~MHz for optimal MOT trapping. We employ a radio-frequency MOT configuration~\cite{norrgard_rfmot_2016} where laser polarizations and the magnetic field gradient are switched at a rate of $0.9$~MHz to remix the magnetic dark states. The laser polarizations are switched between $\sigma^+$ and $\sigma^-$ with high contrast, while the magnetic field gradient is switched by resonantly, sinusoidally modulating the current into the anti-Helmholtz MOT coils, producing a root-mean-square axial gradient of $17$~G$/$cm. Trapping is established when the switching is synchronized in phase.

\begin{figure}[t]
   \centering
   \includegraphics[scale=0.41]{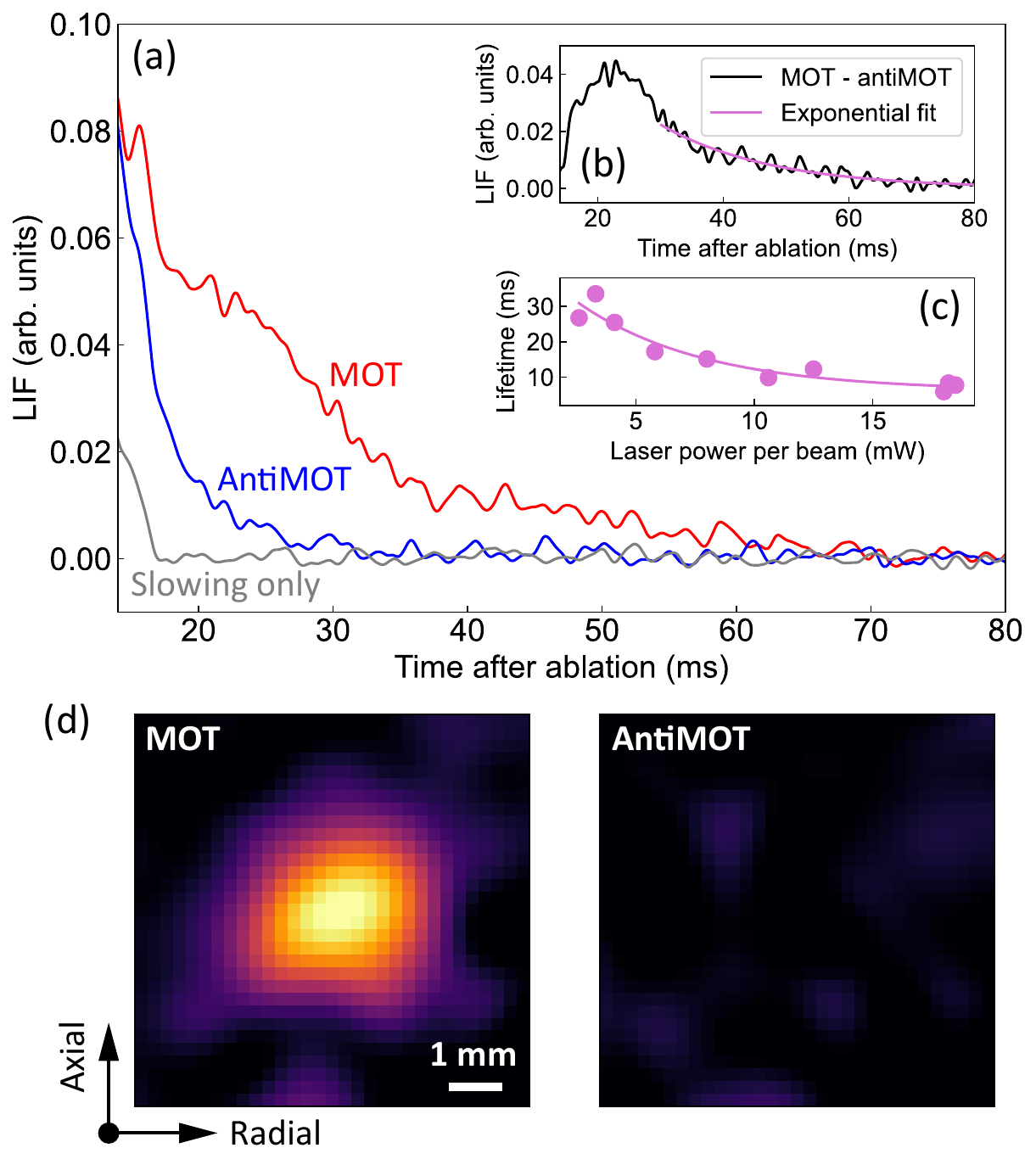}    
   \caption{CaH MOT measurements. (a) LIF detected with a PMT under three configurations: slowing only (gray), MOT (red), and antiMOT (blue). The presence of molecules in the MOT after the molecular beam has fully traversed the region demonstrates trapping. (b) LIF of the antiMOT phase subtracted from that of the MOT phase; $1/e$ lifetime can be extracted from an exponential fit to the tail of the trace. (c) $1/e$ MOT lifetime as a function of the laser power; $\sim$$30$~ms is achieved with a few milliwatts. The solid line is a guide to the eye. (d) Camera images of the MOT and antiMOT integrating from $25$ to $55$~ms after ablation. The images are smoothed with a Gaussian filter of $0.6$~mm standard deviation.}
   \label{fig:MOT}
\end{figure}

The MOT starts to load when the main cycling laser for slowing is turned off. The repumping lasers for slowing are kept on to serve as repumps for the MOT. We detect the $635$~nm LIF corresponding to $B^2\Sigma^+(\nu'=0,\ N'=0,\ +)\rightarrow X^2\Sigma^+(\nu=0,\ N=1,\ -)$ due to the ($\nu=1$) repumping laser. Figure~\hyperref[fig:MOT]{\ref*{fig:MOT}(a)} shows the LIF detected by a PMT under different configurations: slowing only (gray), MOT (red), and antiMOT (blue). The ``antiMOT'' configuration is realized when the laser polarizations are out of phase with respect to the magnetic field gradient, the trace showing the slowed molecules traversing the MOT region. In the MOT configuration, we detect molecules long after the molecular beam passes the region, demonstrating successful magneto-optical trapping of CaH molecules. For Fig.~\hyperref[fig:MOT]{\ref*{fig:MOT}(a)}, the MOT is loaded with $11$~mW of laser power per beam. After $\sim$$10$~ms of MOT loading and thermalization ($25$~ms after ablation), the power is ramped down to $7.5$~mW linearly over $10$~ms to extend the trap lifetime and lower its temperature. To determine the MOT lifetime, we subtract the LIF of the antiMOT phase from that of the MOT phase and fit the tail of the differential LIF to an exponential decay, as shown in Fig.~\hyperref[fig:MOT]{\ref*{fig:MOT}(b)}. We measure the corresponding $1/e$ lifetime at various MOT laser powers and obtain lifetimes up to $\sim$$30$~ms at a few milliwatts [Fig.~\hyperref[fig:MOT]{\ref*{fig:MOT}(c)}]. From the observed trend, we expect even longer lifetimes with lower laser powers, but with the smaller number of trapped molecules, significantly more averages are needed for comparable SNRs.

We directly image the molecules with an electron-multiplying charge-coupled device camera, integrating from $25$ to $55$~ms after ablation for both the MOT and the antiMOT [Fig.~\hyperref[fig:MOT]{\ref*{fig:MOT}(d)}]. The high contrast between the two configurations confirms the successful creation of a CaH MOT. The number of trapped molecules is extracted from the camera's photoelectron counts. After calibrating the photon collection efficiency of the imaging system and measuring photon scattering rates, we estimate a peak number of $230(40)$~trapped molecules. The photon scattering rates are determined by turning off the ($\nu=2$) repumping laser and measuring the $1/e$ lifetime. With the knowledge of the VBR to the ($\nu=2$) state, the photon scattering rate can be inferred. This measurement also enables estimates of loss to higher vibrational states or predissociation of the $B$ states~\cite{sun2023probing}. The vibrational leakage to ($\nu=3$) should not play a significant role in the MOT given the photon budget. The loss is, therefore, predominantly due to predissociation of the $B^2\Sigma^+(\nu'=0,\ N'=0,\ +)$ excited state used for repumping ($\nu=1$). With the measured MOT lifetime of $15.1(2)$~ms and photon scattering rate of $6.1(1.1)\times10^5~\text{s}^{-1}$ at $8$~mW of laser power per beam, the predissociation probability is estimated to be $3.7(7)\times10^{-3}$. This value is larger than our previous measurement, though in better agreement with theoretical predictions~\cite{sun2023probing}, and reduces the photon budget to $\sim$$9\times 10^3$. Given that $\sim$$7\times 10^3$ photons must be scattered for effective laser slowing, $\lesssim$$0.5$ of the initial population can survive this process, limiting the number of trappable molecules.

%%%%%%%%%%%%%%%%%%%%%%%%%%%%%%% MOT characterization %%%%%%%%%%%%%%%%%%%%%%%%%%%%%%%

% \section{MOT characterization}

\begin{figure}[t]
   \centering
   \includegraphics[scale=0.41]{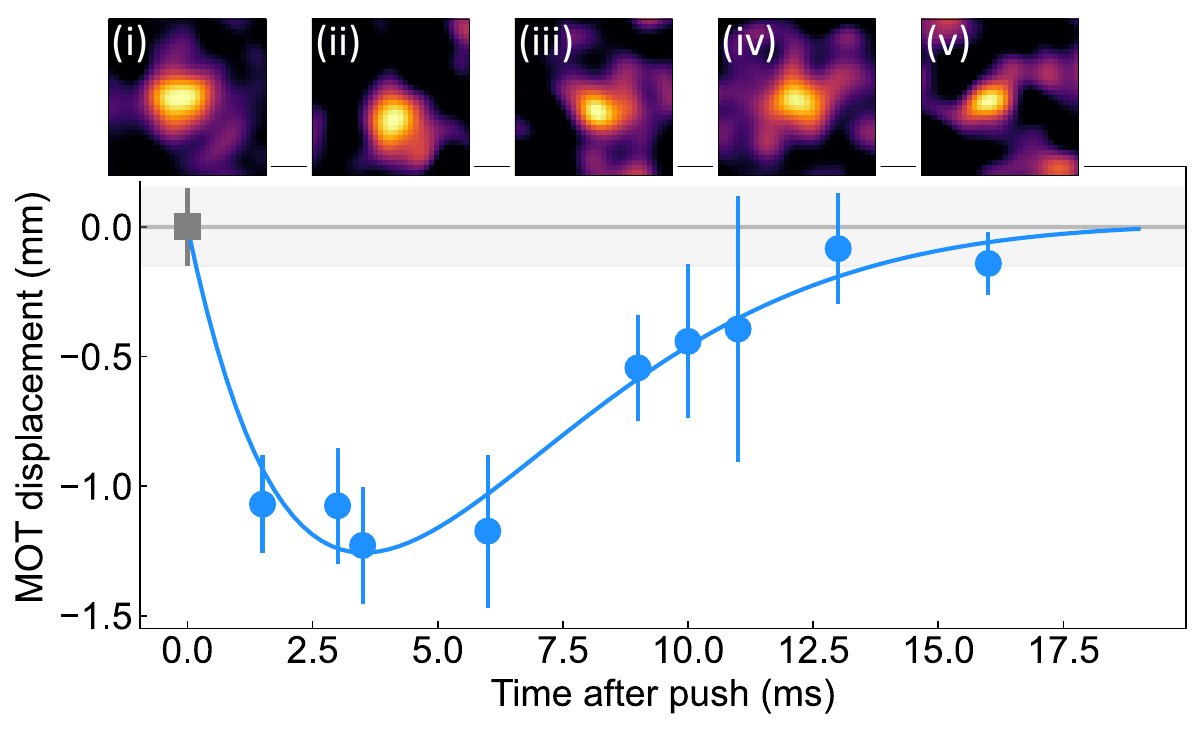}    
   \caption{MOT trapping and cooling force measurement. MOT displacement as a function of time after an applied push by the main cycling laser for slowing. Insets (ii)--(v) are sample MOT images at $3.5$, $9$, $13$, and $16$~ms after the push. Inset (i) shows the unperturbed MOT, with its fitted center position marked by the gray square point and the corresponding shaded region. The images are smoothed with a Gaussian filter of $0.6$~mm standard deviation and normalized to the same scale for better visualization. Error bars represent 1-$\sigma$ uncertainties. The fit of the displacements versus time yields a trapping frequency of $\omega=2\pi\times48(3)$~Hz and a damping constant of $\beta=510(110)$~s$^{-1}$.}
   \label{fig:MOT_push}
\end{figure}

To characterize the MOT, we measure its properties, including the trapping and cooling forces, the size of the trapped molecular cloud, and the temperature. In the low temperature regime, the force $F$ on the trapped molecules can be approximated with a damped harmonic oscillator as $F/m=-\omega^2r-\beta v$, where $\omega$ and $\beta$ are the trapping frequency and damping constant, respectively. These are measured by observing the oscillation of the molecules in the MOT after a push applied by the main cycling laser for slowing. We first load the molecules with $11$~mW of MOT laser power per beam. After thermalization, $25$~ms after ablation, the main cycling laser for slowing is pulsed on for $0.5$~ms. The MOT displacement from equilibrium as a function of time can be described by the expression $r(t)\propto e^{-\beta t/2}\sin[\sqrt{\omega^2-(\beta/2)^2}t]$ and is measured in Fig.~\ref{fig:MOT_push}. Figures~\hyperref[fig:MOT_push]{\ref*{fig:MOT_push}(ii)}--\hyperref[fig:MOT_push]{\ref*{fig:MOT_push}(v)} are sample images showing the MOT at $3.5$, $9$, $13$, and $16$~ms after the push. For each measurement, the molecules are imaged for $2$~ms, and the displacement is extracted from a two-dimensional Gaussian fit. Figure~\hyperref[fig:MOT_push]{\ref*{fig:MOT_push}(i)} shows the unperturbed MOT, with the gray square point and the corresponding shaded region marking its fitted position. The trapping frequency and damping constant are measured to be $\omega=2\pi\times48(3)$~Hz and $\beta=510(110)$~s$^{-1}$, which are comparable to those of other molecular MOTs such as that for calcium monohydroxide (CaOH)~\cite{vilas2022magneto}.

\begin{figure}[t]
   \centering
   \includegraphics[scale=0.41]{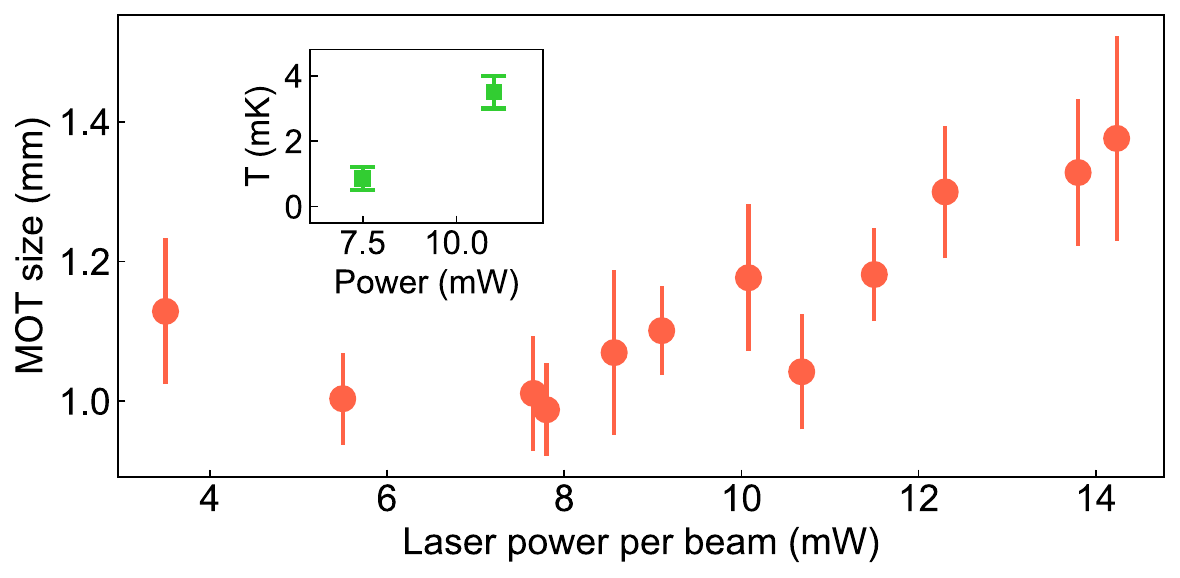}    
   \caption{MOT size and temperature measurements. Geometric mean MOT size as a function of the MOT laser power per beam. The MOT size increases with higher power due to sub-Doppler heating. The inset shows the measured geometric mean temperature of the MOT at two laser powers. Error bars represent 1-$\sigma$ uncertainties.}
   \label{fig:MOT_width}
\end{figure}

Other key MOT characteristics include its size and temperature. We measure the geometric mean MOT size $\sigma_{\text{MOT}}=\sigma_{\parallel}^{1/3}\sigma_{\perp}^{2/3}$ as a function of the MOT laser power (Fig.~\ref{fig:MOT_width}). Here, $\sigma_{\parallel}$ and $\sigma_{\perp}$ are the fitted Gaussian widths of the MOT in the axial and radial directions of the magnetic field gradient, respectively. The camera signal is integrated from $25$ to $55$~ms after ablation, while the MOT laser powers are held constant. We observe an increase in MOT size with higher laser powers, which is expected, since sub-Doppler heating becomes more significant in the regime where the laser intensity exceeds the saturation intensity ($I_{\text{sat}}\approx1.9$~mW$/$cm$^2$). We also measure the MOT temperatures at two characteristic laser powers, $7.5$ and $11$~mW, as shown in the inset of Fig.~\ref{fig:MOT_width}:  after loading and thermalization, we release and then recapture the MOT after a variable wait time. The MOT temperature can be inferred from the time-of-flight expansion, where the evolution of the Gaussian widths in the axial and the radial directions follows $\sigma(t)=\sqrt{\sigma^2(0)+(k_{B}T/m)t^2}$. Here, $k_B$ is the Boltzmann constant. The measured geometric mean temperatures $T_{\text{MOT}}=T_{\parallel}^{1/3}T_{\perp}^{2/3}$ are $0.86(36)$ and $3.5(5)$~mK for the $7.5$ and $11$~mW laser powers.

%%%%%%%%%%%%%%%%%%%%%%%%%%%%%%% Conclusion %%%%%%%%%%%%%%%%%%%%%%%%%%%%%%%

% \section{Conclusion}

In summary, we have demonstrated a three-dimensional MOT of a metal hydride molecule, CaH. The number of trapped molecules is currently limited by the beam source characteristics and by predissociative loss during laser slowing. An increase of the trapped molecule number to $\sim$$10^{3}$ should be possible with higher slowing laser power and chirped laser slowing. Other MOT parameters, including the lifetime, trapping and cooling forces, and temperature, are found to be similar to those of molecular MOTs with comparable masses~\cite{anderegg2017cafmot,Williams_2017,vilas2022magneto}. This work sets the stage for further cooling and trapping experiments with metal hydride molecules, including sub-Doppler cooling~\cite{truppe2017molecules,anderegg_laser_2018,vilas2022magneto}, blue-detuned MOTs~\cite{burau2023bluemot,jorapur2024srfbluemot,li2024cafbluemot,w9qc-rczf,yu_conveyor-belt_2026,pd77-s994}, and optical dipole trapping~\cite{hallas2023caohodt,2h36-myzp}. Optical trapping of hydrogen atoms for precision spectroscopy should become feasible via near-threshold dissociation~\cite{lane2015hydrogen,sun2023probing}. The demonstrated techniques for metal hydride molecules can be extended to deuterides~\cite{dai2024laser}, potentially contributing to isotope shift measurements in the search for physics beyond the standard model~\cite{potvliege2023deuterium}. Furthermore, this new laser cooled and trapped molecular species opens new possibilities at the forefront of quantum chemistry~\cite{liu2022bimolecular,tscherbul2020magnetic}.

%%%%%%%%%%%%%%%%%%%%%%%%%%%%%%% Acknowledgments %%%%%%%%%%%%%%%%%%%%%%%%%%%%%%%

% \section{Acknowledgments}

We thank Y. Lu, Y. Bao, T. K. Langin, Z. Zeng, N. B. Vilas, Z. D. Lasner, and B. Yan for valuable discussions and C. Convery for early experimental contributions.  This work was supported by the ONR Grant No. N00014-25-1-2324, AFOSR MURI Grant No. FA9550-21-1-0069, and the Brown Science Foundation.

%%%%%%%%%%%%%%%%%%%%%%%%%%%%%%% References %%%%%%%%%%%%%%%%%%%%%%%%%%%%%%%

% \bibliographystyle{apsrev4-2}
% \bibliography{references}

\begin{thebibliography}{55}%
\makeatletter
\providecommand \@ifxundefined [1]{%
 \@ifx{#1\undefined}
}%
\providecommand \@ifnum [1]{%
 \ifnum #1\expandafter \@firstoftwo
 \else \expandafter \@secondoftwo
 \fi
}%
\providecommand \@ifx [1]{%
 \ifx #1\expandafter \@firstoftwo
 \else \expandafter \@secondoftwo
 \fi
}%
\providecommand \natexlab [1]{#1}%
\providecommand \enquote  [1]{``#1''}%
\providecommand \bibnamefont  [1]{#1}%
\providecommand \bibfnamefont [1]{#1}%
\providecommand \citenamefont [1]{#1}%
\providecommand \href@noop [0]{\@secondoftwo}%
\providecommand \href [0]{\begingroup \@sanitize@url \@href}%
\providecommand \@href[1]{\@@startlink{#1}\@@href}%
\providecommand \@@href[1]{\endgroup#1\@@endlink}%
\providecommand \@sanitize@url [0]{\catcode `\\12\catcode `\$12\catcode
  `\&12\catcode `\#12\catcode `\^12\catcode `\_12\catcode `\%12\relax}%
\providecommand \@@startlink[1]{}%
\providecommand \@@endlink[0]{}%
\providecommand \url  [0]{\begingroup\@sanitize@url \@url }%
\providecommand \@url [1]{\endgroup\@href {#1}{\urlprefix }}%
\providecommand \urlprefix  [0]{URL }%
\providecommand \Eprint [0]{\href }%
\providecommand \doibase [0]{https://doi.org/}%
\providecommand \selectlanguage [0]{\@gobble}%
\providecommand \bibinfo  [0]{\@secondoftwo}%
\providecommand \bibfield  [0]{\@secondoftwo}%
\providecommand \translation [1]{[#1]}%
\providecommand \BibitemOpen [0]{}%
\providecommand \bibitemStop [0]{}%
\providecommand \bibitemNoStop [0]{.\EOS\space}%
\providecommand \EOS [0]{\spacefactor3000\relax}%
\providecommand \BibitemShut  [1]{\csname bibitem#1\endcsname}%
\let\auto@bib@innerbib\@empty
%</preamble>
\bibitem [{\citenamefont {Holland}\ \emph {et~al.}(2023)\citenamefont
  {Holland}, \citenamefont {Lu},\ and\ \citenamefont
  {Cheuk}}]{holland2023cafentangle}%
  \BibitemOpen
  \bibfield  {author} {\bibinfo {author} {\bibfnamefont {C.~M.}\ \bibnamefont
  {Holland}}, \bibinfo {author} {\bibfnamefont {Y.}~\bibnamefont {Lu}},\ and\
  \bibinfo {author} {\bibfnamefont {L.~W.}\ \bibnamefont {Cheuk}},\ }\bibfield
  {title} {\bibinfo {title} {On-demand entanglement of molecules in a
  reconfigurable optical tweezer array},\ }\href
  {https://doi.org/10.1126/science.adf4272} {\bibfield  {journal} {\bibinfo
  {journal} {Science}\ }\textbf {\bibinfo {volume} {382}},\ \bibinfo {pages}
  {1143} (\bibinfo {year} {2023})}\BibitemShut {NoStop}%
\bibitem [{\citenamefont {Bao}\ \emph {et~al.}(2023)\citenamefont {Bao},
  \citenamefont {Yu}, \citenamefont {Anderegg}, \citenamefont {Chae},
  \citenamefont {Ketterle}, \citenamefont {Ni},\ and\ \citenamefont
  {Doyle}}]{bao2023cafentangle}%
  \BibitemOpen
  \bibfield  {author} {\bibinfo {author} {\bibfnamefont {Y.}~\bibnamefont
  {Bao}}, \bibinfo {author} {\bibfnamefont {S.~S.}\ \bibnamefont {Yu}},
  \bibinfo {author} {\bibfnamefont {L.}~\bibnamefont {Anderegg}}, \bibinfo
  {author} {\bibfnamefont {E.}~\bibnamefont {Chae}}, \bibinfo {author}
  {\bibfnamefont {W.}~\bibnamefont {Ketterle}}, \bibinfo {author}
  {\bibfnamefont {K.-K.}\ \bibnamefont {Ni}},\ and\ \bibinfo {author}
  {\bibfnamefont {J.~M.}\ \bibnamefont {Doyle}},\ }\bibfield  {title} {\bibinfo
  {title} {Dipolar spin-exchange and entanglement between molecules in an
  optical tweezer array},\ }\href {https://doi.org/10.1126/science.adf8999}
  {\bibfield  {journal} {\bibinfo  {journal} {Science}\ }\textbf {\bibinfo
  {volume} {382}},\ \bibinfo {pages} {1138} (\bibinfo {year}
  {2023})}\BibitemShut {NoStop}%
\bibitem [{\citenamefont {Picard}\ \emph {et~al.}(2025)\citenamefont {Picard},
  \citenamefont {Park}, \citenamefont {Patenotte}, \citenamefont
  {Gebretsadkan}, \citenamefont {Wellnitz}, \citenamefont {Rey},\ and\
  \citenamefont {Ni}}]{picard_entanglement_2025}%
  \BibitemOpen
  \bibfield  {author} {\bibinfo {author} {\bibfnamefont {L.~R.~B.}\
  \bibnamefont {Picard}}, \bibinfo {author} {\bibfnamefont {A.~J.}\
  \bibnamefont {Park}}, \bibinfo {author} {\bibfnamefont {G.~E.}\ \bibnamefont
  {Patenotte}}, \bibinfo {author} {\bibfnamefont {S.}~\bibnamefont
  {Gebretsadkan}}, \bibinfo {author} {\bibfnamefont {D.}~\bibnamefont
  {Wellnitz}}, \bibinfo {author} {\bibfnamefont {A.~M.}\ \bibnamefont {Rey}},\
  and\ \bibinfo {author} {\bibfnamefont {K.-K.}\ \bibnamefont {Ni}},\
  }\bibfield  {title} {\bibinfo {title} {Entanglement and {iSWAP} gate between
  molecular qubits},\ }\href {https://doi.org/10.1038/s41586-024-08177-3}
  {\bibfield  {journal} {\bibinfo  {journal} {Nature}\ }\textbf {\bibinfo
  {volume} {637}},\ \bibinfo {pages} {821} (\bibinfo {year}
  {2025})}\BibitemShut {NoStop}%
\bibitem [{\citenamefont {Ruttley}\ \emph {et~al.}(2025)\citenamefont
  {Ruttley}, \citenamefont {Hepworth}, \citenamefont {Guttridge},\ and\
  \citenamefont {Cornish}}]{ruttley_long-lived_2025}%
  \BibitemOpen
  \bibfield  {author} {\bibinfo {author} {\bibfnamefont {D.~K.}\ \bibnamefont
  {Ruttley}}, \bibinfo {author} {\bibfnamefont {T.~R.}\ \bibnamefont
  {Hepworth}}, \bibinfo {author} {\bibfnamefont {A.}~\bibnamefont
  {Guttridge}},\ and\ \bibinfo {author} {\bibfnamefont {S.~L.}\ \bibnamefont
  {Cornish}},\ }\bibfield  {title} {\bibinfo {title} {Long-lived entanglement
  of molecules in magic-wavelength optical tweezers},\ }\href
  {https://doi.org/10.1038/s41586-024-08365-1} {\bibfield  {journal} {\bibinfo
  {journal} {Nature}\ }\textbf {\bibinfo {volume} {637}},\ \bibinfo {pages}
  {827} (\bibinfo {year} {2025})}\BibitemShut {NoStop}%
\bibitem [{\citenamefont {Liu}\ and\ \citenamefont
  {Ni}(2022)}]{liu2022bimolecular}%
  \BibitemOpen
  \bibfield  {author} {\bibinfo {author} {\bibfnamefont {Y.}~\bibnamefont
  {Liu}}\ and\ \bibinfo {author} {\bibfnamefont {K.-K.}\ \bibnamefont {Ni}},\
  }\bibfield  {title} {\bibinfo {title} {Bimolecular chemistry in the ultracold
  regime},\ }\href
  {https://doi.org/https://doi.org/10.1146/annurev-physchem-090419-043244}
  {\bibfield  {journal} {\bibinfo  {journal} {Annu. Rev. Phys. Chem.}\ }\textbf
  {\bibinfo {volume} {73}},\ \bibinfo {pages} {73} (\bibinfo {year}
  {2022})}\BibitemShut {NoStop}%
\bibitem [{\citenamefont {McDonald}\ \emph {et~al.}(2016)\citenamefont
  {McDonald}, \citenamefont {McGuyer}, \citenamefont {Apfelbeck}, \citenamefont
  {Lee}, \citenamefont {Majewska}, \citenamefont {Moszynski},\ and\
  \citenamefont {Zelevinsky}}]{ZelevinskyMcDonaldNature16_Sr2PD}%
  \BibitemOpen
  \bibfield  {author} {\bibinfo {author} {\bibfnamefont {M.}~\bibnamefont
  {McDonald}}, \bibinfo {author} {\bibfnamefont {B.~H.}\ \bibnamefont
  {McGuyer}}, \bibinfo {author} {\bibfnamefont {F.}~\bibnamefont {Apfelbeck}},
  \bibinfo {author} {\bibfnamefont {C.-H.}\ \bibnamefont {Lee}}, \bibinfo
  {author} {\bibfnamefont {I.}~\bibnamefont {Majewska}}, \bibinfo {author}
  {\bibfnamefont {R.}~\bibnamefont {Moszynski}},\ and\ \bibinfo {author}
  {\bibfnamefont {T.}~\bibnamefont {Zelevinsky}},\ }\bibfield  {title}
  {\bibinfo {title} {Photodissociation of ultracold diatomic strontium
  molecules with quantum state control},\ }\href
  {https://doi.org/10.1038/nature18314} {\bibfield  {journal} {\bibinfo
  {journal} {Nature}\ }\textbf {\bibinfo {volume} {534}},\ \bibinfo {pages}
  {122} (\bibinfo {year} {2016})}\BibitemShut {NoStop}%
\bibitem [{\citenamefont {Liu}\ \emph {et~al.}(2025)\citenamefont {Liu},
  \citenamefont {Zhu}, \citenamefont {Luke}, \citenamefont {Babin},
  \citenamefont {Gronowski}, \citenamefont {Ladjimi}, \citenamefont {Tomza},
  \citenamefont {Bohn}, \citenamefont {Tscherbul},\ and\ \citenamefont
  {Ni}}]{liu_hyperfine--rotational_2025}%
  \BibitemOpen
  \bibfield  {author} {\bibinfo {author} {\bibfnamefont {Y.-X.}\ \bibnamefont
  {Liu}}, \bibinfo {author} {\bibfnamefont {L.}~\bibnamefont {Zhu}}, \bibinfo
  {author} {\bibfnamefont {J.}~\bibnamefont {Luke}}, \bibinfo {author}
  {\bibfnamefont {M.~C.}\ \bibnamefont {Babin}}, \bibinfo {author}
  {\bibfnamefont {M.}~\bibnamefont {Gronowski}}, \bibinfo {author}
  {\bibfnamefont {H.}~\bibnamefont {Ladjimi}}, \bibinfo {author} {\bibfnamefont
  {M.}~\bibnamefont {Tomza}}, \bibinfo {author} {\bibfnamefont {J.~L.}\
  \bibnamefont {Bohn}}, \bibinfo {author} {\bibfnamefont {T.~V.}\ \bibnamefont
  {Tscherbul}},\ and\ \bibinfo {author} {\bibfnamefont {K.-K.}\ \bibnamefont
  {Ni}},\ }\bibfield  {title} {\bibinfo {title} {Hyperfine-to-rotational energy
  transfer in ultracold atom–molecule collisions of {Rb} and {KRb}},\ }\href
  {https://doi.org/10.1038/s41557-025-01778-z} {\bibfield  {journal} {\bibinfo
  {journal} {Nat. Chem.}\ }\textbf {\bibinfo {volume} {17}},\ \bibinfo {pages}
  {688} (\bibinfo {year} {2025})}\BibitemShut {NoStop}%
\bibitem [{\citenamefont {Leung}\ \emph {et~al.}(2023)\citenamefont {Leung},
  \citenamefont {Iritani}, \citenamefont {Tiberi}, \citenamefont {Majewska},
  \citenamefont {Borkowski}, \citenamefont {Moszynski},\ and\ \citenamefont
  {Zelevinsky}}]{leung2023molecularclock}%
  \BibitemOpen
  \bibfield  {author} {\bibinfo {author} {\bibfnamefont {K.~H.}\ \bibnamefont
  {Leung}}, \bibinfo {author} {\bibfnamefont {B.}~\bibnamefont {Iritani}},
  \bibinfo {author} {\bibfnamefont {E.}~\bibnamefont {Tiberi}}, \bibinfo
  {author} {\bibfnamefont {I.}~\bibnamefont {Majewska}}, \bibinfo {author}
  {\bibfnamefont {M.}~\bibnamefont {Borkowski}}, \bibinfo {author}
  {\bibfnamefont {R.}~\bibnamefont {Moszynski}},\ and\ \bibinfo {author}
  {\bibfnamefont {T.}~\bibnamefont {Zelevinsky}},\ }\bibfield  {title}
  {\bibinfo {title} {Terahertz vibrational molecular clock with systematic
  uncertainty at the ${10}^{\ensuremath{-}14}$ level},\ }\href
  {https://doi.org/10.1103/PhysRevX.13.011047} {\bibfield  {journal} {\bibinfo
  {journal} {Phys. Rev. X}\ }\textbf {\bibinfo {volume} {13}},\ \bibinfo
  {pages} {011047} (\bibinfo {year} {2023})}\BibitemShut {NoStop}%
\bibitem [{\citenamefont {Anderegg}\ \emph {et~al.}(2023)\citenamefont
  {Anderegg}, \citenamefont {Vilas}, \citenamefont {Hallas}, \citenamefont
  {Robichaud}, \citenamefont {Jadbabaie}, \citenamefont {Doyle},\ and\
  \citenamefont {Hutzler}}]{anderegg2023caohedm}%
  \BibitemOpen
  \bibfield  {author} {\bibinfo {author} {\bibfnamefont {L.}~\bibnamefont
  {Anderegg}}, \bibinfo {author} {\bibfnamefont {N.~B.}\ \bibnamefont {Vilas}},
  \bibinfo {author} {\bibfnamefont {C.}~\bibnamefont {Hallas}}, \bibinfo
  {author} {\bibfnamefont {P.}~\bibnamefont {Robichaud}}, \bibinfo {author}
  {\bibfnamefont {A.}~\bibnamefont {Jadbabaie}}, \bibinfo {author}
  {\bibfnamefont {J.~M.}\ \bibnamefont {Doyle}},\ and\ \bibinfo {author}
  {\bibfnamefont {N.~R.}\ \bibnamefont {Hutzler}},\ }\bibfield  {title}
  {\bibinfo {title} {Quantum control of trapped polyatomic molecules for {eEDM}
  searches},\ }\href {https://doi.org/10.1126/science.adg8155} {\bibfield
  {journal} {\bibinfo  {journal} {Science}\ }\textbf {\bibinfo {volume}
  {382}},\ \bibinfo {pages} {665} (\bibinfo {year} {2023})}\BibitemShut
  {NoStop}%
\bibitem [{\citenamefont {Barry}\ \emph {et~al.}(2014)\citenamefont {Barry},
  \citenamefont {McCarron}, \citenamefont {Norrgard}, \citenamefont
  {Steinecker},\ and\ \citenamefont {DeMille}}]{barry2014magneto}%
  \BibitemOpen
  \bibfield  {author} {\bibinfo {author} {\bibfnamefont {J.~F.}\ \bibnamefont
  {Barry}}, \bibinfo {author} {\bibfnamefont {D.~J.}\ \bibnamefont {McCarron}},
  \bibinfo {author} {\bibfnamefont {E.~B.}\ \bibnamefont {Norrgard}}, \bibinfo
  {author} {\bibfnamefont {M.~H.}\ \bibnamefont {Steinecker}},\ and\ \bibinfo
  {author} {\bibfnamefont {D.}~\bibnamefont {DeMille}},\ }\bibfield  {title}
  {\bibinfo {title} {Magneto-optical trapping of a diatomic molecule},\ }\href
  {https://doi.org/10.1038/nature13634} {\bibfield  {journal} {\bibinfo
  {journal} {Nature}\ }\textbf {\bibinfo {volume} {512}},\ \bibinfo {pages}
  {286} (\bibinfo {year} {2014})}\BibitemShut {NoStop}%
\bibitem [{\citenamefont {Truppe}\ \emph {et~al.}(2017)\citenamefont {Truppe},
  \citenamefont {Williams}, \citenamefont {Hambach}, \citenamefont {Caldwell},
  \citenamefont {Fitch}, \citenamefont {Hinds}, \citenamefont {Sauer},\ and\
  \citenamefont {Tarbutt}}]{truppe2017molecules}%
  \BibitemOpen
  \bibfield  {author} {\bibinfo {author} {\bibfnamefont {S.}~\bibnamefont
  {Truppe}}, \bibinfo {author} {\bibfnamefont {H.~J.}\ \bibnamefont
  {Williams}}, \bibinfo {author} {\bibfnamefont {M.}~\bibnamefont {Hambach}},
  \bibinfo {author} {\bibfnamefont {L.}~\bibnamefont {Caldwell}}, \bibinfo
  {author} {\bibfnamefont {N.~J.}\ \bibnamefont {Fitch}}, \bibinfo {author}
  {\bibfnamefont {E.~A.}\ \bibnamefont {Hinds}}, \bibinfo {author}
  {\bibfnamefont {B.~E.}\ \bibnamefont {Sauer}},\ and\ \bibinfo {author}
  {\bibfnamefont {M.~R.}\ \bibnamefont {Tarbutt}},\ }\bibfield  {title}
  {\bibinfo {title} {{Molecules cooled below the Doppler limit}},\ }\href
  {https://doi.org/10.1038/nphys4241} {\bibfield  {journal} {\bibinfo
  {journal} {Nat. Phys.}\ }\textbf {\bibinfo {volume} {13}},\ \bibinfo {pages}
  {1173} (\bibinfo {year} {2017})}\BibitemShut {NoStop}%
\bibitem [{\citenamefont {Anderegg}\ \emph {et~al.}(2017)\citenamefont
  {Anderegg}, \citenamefont {Augenbraun}, \citenamefont {Chae}, \citenamefont
  {Hemmerling}, \citenamefont {Hutzler}, \citenamefont {Ravi}, \citenamefont
  {Collopy}, \citenamefont {Ye}, \citenamefont {Ketterle},\ and\ \citenamefont
  {Doyle}}]{anderegg2017cafmot}%
  \BibitemOpen
  \bibfield  {author} {\bibinfo {author} {\bibfnamefont {L.}~\bibnamefont
  {Anderegg}}, \bibinfo {author} {\bibfnamefont {B.~L.}\ \bibnamefont
  {Augenbraun}}, \bibinfo {author} {\bibfnamefont {E.}~\bibnamefont {Chae}},
  \bibinfo {author} {\bibfnamefont {B.}~\bibnamefont {Hemmerling}}, \bibinfo
  {author} {\bibfnamefont {N.~R.}\ \bibnamefont {Hutzler}}, \bibinfo {author}
  {\bibfnamefont {A.}~\bibnamefont {Ravi}}, \bibinfo {author} {\bibfnamefont
  {A.}~\bibnamefont {Collopy}}, \bibinfo {author} {\bibfnamefont
  {J.}~\bibnamefont {Ye}}, \bibinfo {author} {\bibfnamefont {W.}~\bibnamefont
  {Ketterle}},\ and\ \bibinfo {author} {\bibfnamefont {J.~M.}\ \bibnamefont
  {Doyle}},\ }\bibfield  {title} {\bibinfo {title} {Radio frequency
  magneto-optical trapping of {CaF} with high density},\ }\href
  {https://doi.org/10.1103/PhysRevLett.119.103201} {\bibfield  {journal}
  {\bibinfo  {journal} {Phys. Rev. Lett.}\ }\textbf {\bibinfo {volume} {119}},\
  \bibinfo {pages} {103201} (\bibinfo {year} {2017})}\BibitemShut {NoStop}%
\bibitem [{\citenamefont {Williams}\ \emph {et~al.}(2017)\citenamefont
  {Williams}, \citenamefont {Truppe}, \citenamefont {Hambach}, \citenamefont
  {Caldwell}, \citenamefont {Fitch}, \citenamefont {Hinds}, \citenamefont
  {Sauer},\ and\ \citenamefont {Tarbutt}}]{Williams_2017}%
  \BibitemOpen
  \bibfield  {author} {\bibinfo {author} {\bibfnamefont {H.~J.}\ \bibnamefont
  {Williams}}, \bibinfo {author} {\bibfnamefont {S.}~\bibnamefont {Truppe}},
  \bibinfo {author} {\bibfnamefont {M.}~\bibnamefont {Hambach}}, \bibinfo
  {author} {\bibfnamefont {L.}~\bibnamefont {Caldwell}}, \bibinfo {author}
  {\bibfnamefont {N.~J.}\ \bibnamefont {Fitch}}, \bibinfo {author}
  {\bibfnamefont {E.~A.}\ \bibnamefont {Hinds}}, \bibinfo {author}
  {\bibfnamefont {B.~E.}\ \bibnamefont {Sauer}},\ and\ \bibinfo {author}
  {\bibfnamefont {M.~R.}\ \bibnamefont {Tarbutt}},\ }\bibfield  {title}
  {\bibinfo {title} {Characteristics of a magneto-optical trap of molecules},\
  }\href {https://doi.org/10.1088/1367-2630/aa8e52} {\bibfield  {journal}
  {\bibinfo  {journal} {New J. Phys.}\ }\textbf {\bibinfo {volume} {19}},\
  \bibinfo {pages} {113035} (\bibinfo {year} {2017})}\BibitemShut {NoStop}%
\bibitem [{\citenamefont {Collopy}\ \emph {et~al.}(2018)\citenamefont
  {Collopy}, \citenamefont {Ding}, \citenamefont {Wu}, \citenamefont
  {Finneran}, \citenamefont {Anderegg}, \citenamefont {Augenbraun},
  \citenamefont {Doyle},\ and\ \citenamefont {Ye}}]{collopy20183d}%
  \BibitemOpen
  \bibfield  {author} {\bibinfo {author} {\bibfnamefont {A.~L.}\ \bibnamefont
  {Collopy}}, \bibinfo {author} {\bibfnamefont {S.}~\bibnamefont {Ding}},
  \bibinfo {author} {\bibfnamefont {Y.}~\bibnamefont {Wu}}, \bibinfo {author}
  {\bibfnamefont {I.~A.}\ \bibnamefont {Finneran}}, \bibinfo {author}
  {\bibfnamefont {L.}~\bibnamefont {Anderegg}}, \bibinfo {author}
  {\bibfnamefont {B.~L.}\ \bibnamefont {Augenbraun}}, \bibinfo {author}
  {\bibfnamefont {J.~M.}\ \bibnamefont {Doyle}},\ and\ \bibinfo {author}
  {\bibfnamefont {J.}~\bibnamefont {Ye}},\ }\bibfield  {title} {\bibinfo
  {title} {{3D magneto-optical trap of yttrium monoxide}},\ }\href
  {https://doi.org/10.1103/PhysRevLett.121.213201} {\bibfield  {journal}
  {\bibinfo  {journal} {Phys. Rev. Lett.}\ }\textbf {\bibinfo {volume} {121}},\
  \bibinfo {pages} {213201} (\bibinfo {year} {2018})}\BibitemShut {NoStop}%
\bibitem [{\citenamefont {Vilas}\ \emph {et~al.}(2022)\citenamefont {Vilas},
  \citenamefont {Hallas}, \citenamefont {Anderegg}, \citenamefont {Robichaud},
  \citenamefont {Winnicki}, \citenamefont {Mitra},\ and\ \citenamefont
  {Doyle}}]{vilas2022magneto}%
  \BibitemOpen
  \bibfield  {author} {\bibinfo {author} {\bibfnamefont {N.~B.}\ \bibnamefont
  {Vilas}}, \bibinfo {author} {\bibfnamefont {C.}~\bibnamefont {Hallas}},
  \bibinfo {author} {\bibfnamefont {L.}~\bibnamefont {Anderegg}}, \bibinfo
  {author} {\bibfnamefont {P.}~\bibnamefont {Robichaud}}, \bibinfo {author}
  {\bibfnamefont {A.}~\bibnamefont {Winnicki}}, \bibinfo {author}
  {\bibfnamefont {D.}~\bibnamefont {Mitra}},\ and\ \bibinfo {author}
  {\bibfnamefont {J.~M.}\ \bibnamefont {Doyle}},\ }\bibfield  {title} {\bibinfo
  {title} {{Magneto-optical trapping and sub-Doppler cooling of a polyatomic
  molecule}},\ }\href {https://doi.org/10.1038/s41586-022-04620-5} {\bibfield
  {journal} {\bibinfo  {journal} {Nature}\ }\textbf {\bibinfo {volume} {606}},\
  \bibinfo {pages} {70} (\bibinfo {year} {2022})}\BibitemShut {NoStop}%
\bibitem [{\citenamefont {Zeng}\ \emph {et~al.}(2024)\citenamefont {Zeng},
  \citenamefont {Deng}, \citenamefont {Yang},\ and\ \citenamefont
  {Yan}}]{zeng2024three}%
  \BibitemOpen
  \bibfield  {author} {\bibinfo {author} {\bibfnamefont {Z.}~\bibnamefont
  {Zeng}}, \bibinfo {author} {\bibfnamefont {S.}~\bibnamefont {Deng}}, \bibinfo
  {author} {\bibfnamefont {S.}~\bibnamefont {Yang}},\ and\ \bibinfo {author}
  {\bibfnamefont {B.}~\bibnamefont {Yan}},\ }\bibfield  {title} {\bibinfo
  {title} {Three-dimensional magneto-optical trapping of barium monofluoride},\
  }\href {https://doi.org/10.1103/PhysRevLett.133.143404} {\bibfield  {journal}
  {\bibinfo  {journal} {Phys. Rev. Lett.}\ }\textbf {\bibinfo {volume} {133}},\
  \bibinfo {pages} {143404} (\bibinfo {year} {2024})}\BibitemShut {NoStop}%
\bibitem [{\citenamefont {Lasner}\ \emph {et~al.}(2025)\citenamefont {Lasner},
  \citenamefont {Frenett}, \citenamefont {Sawaoka}, \citenamefont {Anderegg},
  \citenamefont {Augenbraun}, \citenamefont {Lampson}, \citenamefont {Li},
  \citenamefont {Lunstad}, \citenamefont {Mango}, \citenamefont {Nasir},
  \citenamefont {Ono}, \citenamefont {Sakamoto},\ and\ \citenamefont
  {Doyle}}]{lasner2025magneto}%
  \BibitemOpen
  \bibfield  {author} {\bibinfo {author} {\bibfnamefont {Z.~D.}\ \bibnamefont
  {Lasner}}, \bibinfo {author} {\bibfnamefont {A.}~\bibnamefont {Frenett}},
  \bibinfo {author} {\bibfnamefont {H.}~\bibnamefont {Sawaoka}}, \bibinfo
  {author} {\bibfnamefont {L.}~\bibnamefont {Anderegg}}, \bibinfo {author}
  {\bibfnamefont {B.}~\bibnamefont {Augenbraun}}, \bibinfo {author}
  {\bibfnamefont {H.}~\bibnamefont {Lampson}}, \bibinfo {author} {\bibfnamefont
  {M.}~\bibnamefont {Li}}, \bibinfo {author} {\bibfnamefont {A.}~\bibnamefont
  {Lunstad}}, \bibinfo {author} {\bibfnamefont {J.}~\bibnamefont {Mango}},
  \bibinfo {author} {\bibfnamefont {A.}~\bibnamefont {Nasir}}, \bibinfo
  {author} {\bibfnamefont {T.}~\bibnamefont {Ono}}, \bibinfo {author}
  {\bibfnamefont {T.}~\bibnamefont {Sakamoto}},\ and\ \bibinfo {author}
  {\bibfnamefont {J.~M.}\ \bibnamefont {Doyle}},\ }\bibfield  {title} {\bibinfo
  {title} {Magneto-optical trapping of a heavy polyatomic molecule for
  precision measurement},\ }\href
  {https://doi.org/10.1103/PhysRevLett.134.083401} {\bibfield  {journal}
  {\bibinfo  {journal} {Phys. Rev. Lett.}\ }\textbf {\bibinfo {volume} {134}},\
  \bibinfo {pages} {083401} (\bibinfo {year} {2025})}\BibitemShut {NoStop}%
\bibitem [{\citenamefont {Padilla-Castillo}\ \emph {et~al.}(2025)\citenamefont
  {Padilla-Castillo}, \citenamefont {Cai}, \citenamefont {Agarwal},
  \citenamefont {Kukreja}, \citenamefont {Thomas}, \citenamefont {Sartakov},
  \citenamefont {Truppe}, \citenamefont {Meijer},\ and\ \citenamefont
  {Wright}}]{alfmot2025}%
  \BibitemOpen
  \bibfield  {author} {\bibinfo {author} {\bibfnamefont {J.~E.}\ \bibnamefont
  {Padilla-Castillo}}, \bibinfo {author} {\bibfnamefont {J.}~\bibnamefont
  {Cai}}, \bibinfo {author} {\bibfnamefont {P.}~\bibnamefont {Agarwal}},
  \bibinfo {author} {\bibfnamefont {P.}~\bibnamefont {Kukreja}}, \bibinfo
  {author} {\bibfnamefont {R.}~\bibnamefont {Thomas}}, \bibinfo {author}
  {\bibfnamefont {B.~G.}\ \bibnamefont {Sartakov}}, \bibinfo {author}
  {\bibfnamefont {S.}~\bibnamefont {Truppe}}, \bibinfo {author} {\bibfnamefont
  {G.}~\bibnamefont {Meijer}},\ and\ \bibinfo {author} {\bibfnamefont {S.~C.}\
  \bibnamefont {Wright}},\ }\bibfield  {title} {\bibinfo {title}
  {Magneto-optical trapping of aluminum monofluoride},\ }\href
  {https://doi.org/10.1103/ksnd-9fyf} {\bibfield  {journal} {\bibinfo
  {journal} {Phys. Rev. Lett.}\ }\textbf {\bibinfo {volume} {135}},\ \bibinfo
  {pages} {243401} (\bibinfo {year} {2025})}\BibitemShut {NoStop}%
\bibitem [{\citenamefont {Burau}\ \emph {et~al.}(2023)\citenamefont {Burau},
  \citenamefont {Aggarwal}, \citenamefont {Mehling},\ and\ \citenamefont
  {Ye}}]{burau2023bluemot}%
  \BibitemOpen
  \bibfield  {author} {\bibinfo {author} {\bibfnamefont {J.~J.}\ \bibnamefont
  {Burau}}, \bibinfo {author} {\bibfnamefont {P.}~\bibnamefont {Aggarwal}},
  \bibinfo {author} {\bibfnamefont {K.}~\bibnamefont {Mehling}},\ and\ \bibinfo
  {author} {\bibfnamefont {J.}~\bibnamefont {Ye}},\ }\bibfield  {title}
  {\bibinfo {title} {Blue-detuned magneto-optical trap of molecules},\ }\href
  {https://doi.org/10.1103/PhysRevLett.130.193401} {\bibfield  {journal}
  {\bibinfo  {journal} {Phys. Rev. Lett.}\ }\textbf {\bibinfo {volume} {130}},\
  \bibinfo {pages} {193401} (\bibinfo {year} {2023})}\BibitemShut {NoStop}%
\bibitem [{\citenamefont {Li}\ \emph {et~al.}(2025)\citenamefont {Li},
  \citenamefont {Hallas},\ and\ \citenamefont {Doyle}}]{li_conveyor-belt_2025}%
  \BibitemOpen
  \bibfield  {author} {\bibinfo {author} {\bibfnamefont {G.~K.}\ \bibnamefont
  {Li}}, \bibinfo {author} {\bibfnamefont {C.}~\bibnamefont {Hallas}},\ and\
  \bibinfo {author} {\bibfnamefont {J.~M.}\ \bibnamefont {Doyle}},\ }\bibfield
  {title} {\bibinfo {title} {Conveyor-belt magneto-optical trapping of
  molecules},\ }\href {https://doi.org/10.1088/1367-2630/adc032} {\bibfield
  {journal} {\bibinfo  {journal} {New J. Phys.}\ }\textbf {\bibinfo {volume}
  {27}},\ \bibinfo {pages} {043002} (\bibinfo {year} {2025})}\BibitemShut
  {NoStop}%
\bibitem [{\citenamefont {Frum}\ \emph {et~al.}(1993)\citenamefont {Frum},
  \citenamefont {Oh}, \citenamefont {Cohen},\ and\ \citenamefont
  {Pickett}}]{Frum_rotational_spectra_1993}%
  \BibitemOpen
  \bibfield  {author} {\bibinfo {author} {\bibfnamefont {C.~I.}\ \bibnamefont
  {Frum}}, \bibinfo {author} {\bibfnamefont {J.~J.}\ \bibnamefont {Oh}},
  \bibinfo {author} {\bibfnamefont {E.~A.}\ \bibnamefont {Cohen}},\ and\
  \bibinfo {author} {\bibfnamefont {H.~M.}\ \bibnamefont {Pickett}},\
  }\bibfield  {title} {\bibinfo {title} {Rotational spectra of the
  ${X}^2{\Sigma}^+$ states of {CaH} and {CaD}},\ }\href
  {https://adsabs.harvard.edu/full/1993ApJ...408L..61F} {\bibfield  {journal}
  {\bibinfo  {journal} {Astrophys. J.}\ }\textbf {\bibinfo {volume} {408}},\
  \bibinfo {pages} {L61} (\bibinfo {year} {1993})}\BibitemShut {NoStop}%
\bibitem [{\citenamefont {Alavi}\ and\ \citenamefont
  {Shayesteh}(2018)}]{Alavi2018einsteincoeff}%
  \BibitemOpen
  \bibfield  {author} {\bibinfo {author} {\bibfnamefont {S.~F.}\ \bibnamefont
  {Alavi}}\ and\ \bibinfo {author} {\bibfnamefont {A.}~\bibnamefont
  {Shayesteh}},\ }\bibfield  {title} {\bibinfo {title} {Einstein {A}
  coefficients for rovibronic lines of the ${A}^2 {\Pi} \rightarrow {X}^2
  {\Sigma}^+$ and ${B}^2 {\Sigma}^+ \rightarrow {X}^2 {\Sigma}^+$ transitions
  of {CaH} and {CaD}},\ }\href {https://doi.org/10.1093/MNRAS/STX2681}
  {\bibfield  {journal} {\bibinfo  {journal} {Mon. Not. R. Astron. Soc.}\
  }\textbf {\bibinfo {volume} {474}},\ \bibinfo {pages} {2} (\bibinfo {year}
  {2018})}\BibitemShut {NoStop}%
\bibitem [{\citenamefont {Tscherbul}\ and\ \citenamefont
  {K{\l}os}(2020)}]{tscherbul2020magnetic}%
  \BibitemOpen
  \bibfield  {author} {\bibinfo {author} {\bibfnamefont {T.~V.}\ \bibnamefont
  {Tscherbul}}\ and\ \bibinfo {author} {\bibfnamefont {J.}~\bibnamefont
  {K{\l}os}},\ }\bibfield  {title} {\bibinfo {title} {Magnetic tuning of
  ultracold barrierless chemical reactions},\ }\href
  {https://doi.org/10.1103/PhysRevResearch.2.013117} {\bibfield  {journal}
  {\bibinfo  {journal} {Phys. Rev. Res.}\ }\textbf {\bibinfo {volume} {2}},\
  \bibinfo {pages} {013117} (\bibinfo {year} {2020})}\BibitemShut {NoStop}%
\bibitem [{\citenamefont {Lane}(2015)}]{lane2015hydrogen}%
  \BibitemOpen
  \bibfield  {author} {\bibinfo {author} {\bibfnamefont {I.~C.}\ \bibnamefont
  {Lane}},\ }\bibfield  {title} {\bibinfo {title} {{Production of ultracold
  hydrogen and deuterium via Doppler-cooled Feshbach molecules}},\ }\href
  {https://doi.org/10.1103/PhysRevA.92.022511} {\bibfield  {journal} {\bibinfo
  {journal} {Phys. Rev. A}\ }\textbf {\bibinfo {volume} {92}},\ \bibinfo
  {pages} {022511} (\bibinfo {year} {2015})}\BibitemShut {NoStop}%
\bibitem [{\citenamefont {V{\'a}zquez-Carson}\ \emph
  {et~al.}(2022)\citenamefont {V{\'a}zquez-Carson}, \citenamefont {Sun},
  \citenamefont {Dai}, \citenamefont {Mitra},\ and\ \citenamefont
  {Zelevinsky}}]{vazquez2022direct}%
  \BibitemOpen
  \bibfield  {author} {\bibinfo {author} {\bibfnamefont {S.~F.}\ \bibnamefont
  {V{\'a}zquez-Carson}}, \bibinfo {author} {\bibfnamefont {Q.}~\bibnamefont
  {Sun}}, \bibinfo {author} {\bibfnamefont {J.}~\bibnamefont {Dai}}, \bibinfo
  {author} {\bibfnamefont {D.}~\bibnamefont {Mitra}},\ and\ \bibinfo {author}
  {\bibfnamefont {T.}~\bibnamefont {Zelevinsky}},\ }\bibfield  {title}
  {\bibinfo {title} {Direct laser cooling of calcium monohydride molecules},\
  }\href {https://doi.org/10.1088/1367-2630/ac806c} {\bibfield  {journal}
  {\bibinfo  {journal} {New J. Phys.}\ }\textbf {\bibinfo {volume} {24}},\
  \bibinfo {pages} {083006} (\bibinfo {year} {2022})}\BibitemShut {NoStop}%
\bibitem [{\citenamefont {Sun}\ \emph {et~al.}(2023)\citenamefont {Sun},
  \citenamefont {Dickerson}, \citenamefont {Dai}, \citenamefont {Pope},
  \citenamefont {Cheng}, \citenamefont {Neuhauser}, \citenamefont
  {Alexandrova}, \citenamefont {Mitra},\ and\ \citenamefont
  {Zelevinsky}}]{sun2023probing}%
  \BibitemOpen
  \bibfield  {author} {\bibinfo {author} {\bibfnamefont {Q.}~\bibnamefont
  {Sun}}, \bibinfo {author} {\bibfnamefont {C.~E.}\ \bibnamefont {Dickerson}},
  \bibinfo {author} {\bibfnamefont {J.}~\bibnamefont {Dai}}, \bibinfo {author}
  {\bibfnamefont {I.~M.}\ \bibnamefont {Pope}}, \bibinfo {author}
  {\bibfnamefont {L.}~\bibnamefont {Cheng}}, \bibinfo {author} {\bibfnamefont
  {D.}~\bibnamefont {Neuhauser}}, \bibinfo {author} {\bibfnamefont {A.~N.}\
  \bibnamefont {Alexandrova}}, \bibinfo {author} {\bibfnamefont
  {D.}~\bibnamefont {Mitra}},\ and\ \bibinfo {author} {\bibfnamefont
  {T.}~\bibnamefont {Zelevinsky}},\ }\bibfield  {title} {\bibinfo {title}
  {Probing the limits of optical cycling in a predissociative diatomic
  molecule},\ }\href {https://doi.org/10.1103/PhysRevResearch.5.043070}
  {\bibfield  {journal} {\bibinfo  {journal} {Phys. Rev. Res.}\ }\textbf
  {\bibinfo {volume} {5}},\ \bibinfo {pages} {043070} (\bibinfo {year}
  {2023})}\BibitemShut {NoStop}%
\bibitem [{\citenamefont {Tiesinga}\ \emph {et~al.}(2021)\citenamefont
  {Tiesinga}, \citenamefont {Mohr}, \citenamefont {Newell},\ and\ \citenamefont
  {Taylor}}]{tiesinga2021codata}%
  \BibitemOpen
  \bibfield  {author} {\bibinfo {author} {\bibfnamefont {E.}~\bibnamefont
  {Tiesinga}}, \bibinfo {author} {\bibfnamefont {P.~J.}\ \bibnamefont {Mohr}},
  \bibinfo {author} {\bibfnamefont {D.~B.}\ \bibnamefont {Newell}},\ and\
  \bibinfo {author} {\bibfnamefont {B.~N.}\ \bibnamefont {Taylor}},\ }\bibfield
   {title} {\bibinfo {title} {{CODATA recommended values of the fundamental
  physical constants: 2018}},\ }\href
  {https://doi.org/10.1103/RevModPhys.93.025010} {\bibfield  {journal}
  {\bibinfo  {journal} {Rev. Mod. Phys.}\ }\textbf {\bibinfo {volume} {93}},\
  \bibinfo {pages} {025010} (\bibinfo {year} {2021})}\BibitemShut {NoStop}%
\bibitem [{\citenamefont {Biraben}(2009)}]{biraben_spectroscopy_2009}%
  \BibitemOpen
  \bibfield  {author} {\bibinfo {author} {\bibfnamefont {F.}~\bibnamefont
  {Biraben}},\ }\bibfield  {title} {\bibinfo {title} {Spectroscopy of atomic
  hydrogen},\ }\href {https://doi.org/10.1140/epjst/e2009-01045-3} {\bibfield
  {journal} {\bibinfo  {journal} {Eur. Phys. J. Spec. Top.}\ }\textbf {\bibinfo
  {volume} {172}},\ \bibinfo {pages} {109} (\bibinfo {year}
  {2009})}\BibitemShut {NoStop}%
\bibitem [{\citenamefont {Cesar}\ \emph {et~al.}(1996)\citenamefont {Cesar},
  \citenamefont {Fried}, \citenamefont {Killian}, \citenamefont {Polcyn},
  \citenamefont {Sandberg}, \citenamefont {Yu}, \citenamefont {Greytak},
  \citenamefont {Kleppner},\ and\ \citenamefont {Doyle}}]{cesar1996two}%
  \BibitemOpen
  \bibfield  {author} {\bibinfo {author} {\bibfnamefont {C.~L.}\ \bibnamefont
  {Cesar}}, \bibinfo {author} {\bibfnamefont {D.~G.}\ \bibnamefont {Fried}},
  \bibinfo {author} {\bibfnamefont {T.~C.}\ \bibnamefont {Killian}}, \bibinfo
  {author} {\bibfnamefont {A.~D.}\ \bibnamefont {Polcyn}}, \bibinfo {author}
  {\bibfnamefont {J.~C.}\ \bibnamefont {Sandberg}}, \bibinfo {author}
  {\bibfnamefont {I.~A.}\ \bibnamefont {Yu}}, \bibinfo {author} {\bibfnamefont
  {T.~J.}\ \bibnamefont {Greytak}}, \bibinfo {author} {\bibfnamefont
  {D.}~\bibnamefont {Kleppner}},\ and\ \bibinfo {author} {\bibfnamefont
  {J.~M.}\ \bibnamefont {Doyle}},\ }\bibfield  {title} {\bibinfo {title}
  {Two-photon spectroscopy of trapped atomic hydrogen},\ }\href
  {https://doi.org/10.1103/PhysRevLett.77.255} {\bibfield  {journal} {\bibinfo
  {journal} {Phys. Rev. Lett.}\ }\textbf {\bibinfo {volume} {77}},\ \bibinfo
  {pages} {255} (\bibinfo {year} {1996})}\BibitemShut {NoStop}%
\bibitem [{\citenamefont {Parthey}\ \emph {et~al.}(2011)\citenamefont
  {Parthey}, \citenamefont {Matveev}, \citenamefont {Alnis}, \citenamefont
  {Bernhardt}, \citenamefont {Beyer}, \citenamefont {Holzwarth}, \citenamefont
  {Maistrou}, \citenamefont {Pohl}, \citenamefont {Predehl}, \citenamefont
  {Udem}, \citenamefont {Wilken}, \citenamefont {Kolachevsky}, \citenamefont
  {Abgrall}, \citenamefont {Rovera}, \citenamefont {Salomon}, \citenamefont
  {Laurent},\ and\ \citenamefont {H\"ansch}}]{parthey2011improved}%
  \BibitemOpen
  \bibfield  {author} {\bibinfo {author} {\bibfnamefont {C.~G.}\ \bibnamefont
  {Parthey}}, \bibinfo {author} {\bibfnamefont {A.}~\bibnamefont {Matveev}},
  \bibinfo {author} {\bibfnamefont {J.}~\bibnamefont {Alnis}}, \bibinfo
  {author} {\bibfnamefont {B.}~\bibnamefont {Bernhardt}}, \bibinfo {author}
  {\bibfnamefont {A.}~\bibnamefont {Beyer}}, \bibinfo {author} {\bibfnamefont
  {R.}~\bibnamefont {Holzwarth}}, \bibinfo {author} {\bibfnamefont
  {A.}~\bibnamefont {Maistrou}}, \bibinfo {author} {\bibfnamefont
  {R.}~\bibnamefont {Pohl}}, \bibinfo {author} {\bibfnamefont {K.}~\bibnamefont
  {Predehl}}, \bibinfo {author} {\bibfnamefont {T.}~\bibnamefont {Udem}},
  \bibinfo {author} {\bibfnamefont {T.}~\bibnamefont {Wilken}}, \bibinfo
  {author} {\bibfnamefont {N.}~\bibnamefont {Kolachevsky}}, \bibinfo {author}
  {\bibfnamefont {M.}~\bibnamefont {Abgrall}}, \bibinfo {author} {\bibfnamefont
  {D.}~\bibnamefont {Rovera}}, \bibinfo {author} {\bibfnamefont
  {C.}~\bibnamefont {Salomon}}, \bibinfo {author} {\bibfnamefont
  {P.}~\bibnamefont {Laurent}},\ and\ \bibinfo {author} {\bibfnamefont {T.~W.}\
  \bibnamefont {H\"ansch}},\ }\bibfield  {title} {\bibinfo {title} {{Improved
  measurement of the hydrogen $1S-2S$ transition frequency}},\ }\href
  {https://doi.org/10.1103/PhysRevLett.107.203001} {\bibfield  {journal}
  {\bibinfo  {journal} {Phys. Rev. Lett.}\ }\textbf {\bibinfo {volume} {107}},\
  \bibinfo {pages} {203001} (\bibinfo {year} {2011})}\BibitemShut {NoStop}%
\bibitem [{\citenamefont {Beyer}\ \emph {et~al.}(2017)\citenamefont {Beyer},
  \citenamefont {Maisenbacher}, \citenamefont {Matveev}, \citenamefont {Pohl},
  \citenamefont {Khabarova}, \citenamefont {Grinin}, \citenamefont {Lamour},
  \citenamefont {Yost}, \citenamefont {Hänsch}, \citenamefont {Kolachevsky},\
  and\ \citenamefont {Udem}}]{beyer2017rydberg}%
  \BibitemOpen
  \bibfield  {author} {\bibinfo {author} {\bibfnamefont {A.}~\bibnamefont
  {Beyer}}, \bibinfo {author} {\bibfnamefont {L.}~\bibnamefont {Maisenbacher}},
  \bibinfo {author} {\bibfnamefont {A.}~\bibnamefont {Matveev}}, \bibinfo
  {author} {\bibfnamefont {R.}~\bibnamefont {Pohl}}, \bibinfo {author}
  {\bibfnamefont {K.}~\bibnamefont {Khabarova}}, \bibinfo {author}
  {\bibfnamefont {A.}~\bibnamefont {Grinin}}, \bibinfo {author} {\bibfnamefont
  {T.}~\bibnamefont {Lamour}}, \bibinfo {author} {\bibfnamefont {D.~C.}\
  \bibnamefont {Yost}}, \bibinfo {author} {\bibfnamefont {T.~W.}\ \bibnamefont
  {Hänsch}}, \bibinfo {author} {\bibfnamefont {N.}~\bibnamefont
  {Kolachevsky}},\ and\ \bibinfo {author} {\bibfnamefont {T.}~\bibnamefont
  {Udem}},\ }\bibfield  {title} {\bibinfo {title} {{The Rydberg constant and
  proton size from atomic hydrogen}},\ }\href
  {https://doi.org/10.1126/science.aah6677} {\bibfield  {journal} {\bibinfo
  {journal} {Science}\ }\textbf {\bibinfo {volume} {358}},\ \bibinfo {pages}
  {79} (\bibinfo {year} {2017})}\BibitemShut {NoStop}%
\bibitem [{\citenamefont {Bezginov}\ \emph {et~al.}(2019)\citenamefont
  {Bezginov}, \citenamefont {Valdez}, \citenamefont {Horbatsch}, \citenamefont
  {Marsman}, \citenamefont {Vutha},\ and\ \citenamefont
  {Hessels}}]{bezginov2019lamb}%
  \BibitemOpen
  \bibfield  {author} {\bibinfo {author} {\bibfnamefont {N.}~\bibnamefont
  {Bezginov}}, \bibinfo {author} {\bibfnamefont {T.}~\bibnamefont {Valdez}},
  \bibinfo {author} {\bibfnamefont {M.}~\bibnamefont {Horbatsch}}, \bibinfo
  {author} {\bibfnamefont {A.}~\bibnamefont {Marsman}}, \bibinfo {author}
  {\bibfnamefont {A.~C.}\ \bibnamefont {Vutha}},\ and\ \bibinfo {author}
  {\bibfnamefont {E.~A.}\ \bibnamefont {Hessels}},\ }\bibfield  {title}
  {\bibinfo {title} {{A measurement of the atomic hydrogen Lamb shift and the
  proton charge radius}},\ }\href {https://doi.org/10.1126/science.aau7807}
  {\bibfield  {journal} {\bibinfo  {journal} {Science}\ }\textbf {\bibinfo
  {volume} {365}},\ \bibinfo {pages} {1007} (\bibinfo {year}
  {2019})}\BibitemShut {NoStop}%
\bibitem [{\citenamefont {Grinin}\ \emph {et~al.}(2020)\citenamefont {Grinin},
  \citenamefont {Matveev}, \citenamefont {Yost}, \citenamefont {Maisenbacher},
  \citenamefont {Wirthl}, \citenamefont {Pohl}, \citenamefont {H{\"a}nsch},\
  and\ \citenamefont {Udem}}]{grinin2020two}%
  \BibitemOpen
  \bibfield  {author} {\bibinfo {author} {\bibfnamefont {A.}~\bibnamefont
  {Grinin}}, \bibinfo {author} {\bibfnamefont {A.}~\bibnamefont {Matveev}},
  \bibinfo {author} {\bibfnamefont {D.~C.}\ \bibnamefont {Yost}}, \bibinfo
  {author} {\bibfnamefont {L.}~\bibnamefont {Maisenbacher}}, \bibinfo {author}
  {\bibfnamefont {V.}~\bibnamefont {Wirthl}}, \bibinfo {author} {\bibfnamefont
  {R.}~\bibnamefont {Pohl}}, \bibinfo {author} {\bibfnamefont {T.~W.}\
  \bibnamefont {H{\"a}nsch}},\ and\ \bibinfo {author} {\bibfnamefont
  {T.}~\bibnamefont {Udem}},\ }\bibfield  {title} {\bibinfo {title} {Two-photon
  frequency comb spectroscopy of atomic hydrogen},\ }\href
  {https://doi.org/10.1126/science.abc7776} {\bibfield  {journal} {\bibinfo
  {journal} {Science}\ }\textbf {\bibinfo {volume} {370}},\ \bibinfo {pages}
  {1061} (\bibinfo {year} {2020})}\BibitemShut {NoStop}%
\bibitem [{\citenamefont {Brandt}\ \emph {et~al.}(2022)\citenamefont {Brandt},
  \citenamefont {Cooper}, \citenamefont {Rasor}, \citenamefont {Burkley},
  \citenamefont {Matveev},\ and\ \citenamefont {Yost}}]{brandt2022measurement}%
  \BibitemOpen
  \bibfield  {author} {\bibinfo {author} {\bibfnamefont {A.~D.}\ \bibnamefont
  {Brandt}}, \bibinfo {author} {\bibfnamefont {S.~F.}\ \bibnamefont {Cooper}},
  \bibinfo {author} {\bibfnamefont {C.}~\bibnamefont {Rasor}}, \bibinfo
  {author} {\bibfnamefont {Z.}~\bibnamefont {Burkley}}, \bibinfo {author}
  {\bibfnamefont {A.}~\bibnamefont {Matveev}},\ and\ \bibinfo {author}
  {\bibfnamefont {D.~C.}\ \bibnamefont {Yost}},\ }\bibfield  {title} {\bibinfo
  {title} {{Measurement of the
  $2{\mathrm{S}}_{1/2}\ensuremath{-}8{\mathrm{D}}_{5/2}$ transition in
  hydrogen}},\ }\href {https://doi.org/10.1103/PhysRevLett.128.023001}
  {\bibfield  {journal} {\bibinfo  {journal} {Phys. Rev. Lett.}\ }\textbf
  {\bibinfo {volume} {128}},\ \bibinfo {pages} {023001} (\bibinfo {year}
  {2022})}\BibitemShut {NoStop}%
\bibitem [{\citenamefont {Fried}\ \emph {et~al.}(1998)\citenamefont {Fried},
  \citenamefont {Killian}, \citenamefont {Willmann}, \citenamefont {Landhuis},
  \citenamefont {Moss}, \citenamefont {Kleppner},\ and\ \citenamefont
  {Greytak}}]{fried1998bose}%
  \BibitemOpen
  \bibfield  {author} {\bibinfo {author} {\bibfnamefont {D.~G.}\ \bibnamefont
  {Fried}}, \bibinfo {author} {\bibfnamefont {T.~C.}\ \bibnamefont {Killian}},
  \bibinfo {author} {\bibfnamefont {L.}~\bibnamefont {Willmann}}, \bibinfo
  {author} {\bibfnamefont {D.}~\bibnamefont {Landhuis}}, \bibinfo {author}
  {\bibfnamefont {S.~C.}\ \bibnamefont {Moss}}, \bibinfo {author}
  {\bibfnamefont {D.}~\bibnamefont {Kleppner}},\ and\ \bibinfo {author}
  {\bibfnamefont {T.~J.}\ \bibnamefont {Greytak}},\ }\bibfield  {title}
  {\bibinfo {title} {{Bose-Einstein condensation of atomic hydrogen}},\ }\href
  {https://doi.org/10.1103/PhysRevLett.81.3811} {\bibfield  {journal} {\bibinfo
   {journal} {Phys. Rev. Lett.}\ }\textbf {\bibinfo {volume} {81}},\ \bibinfo
  {pages} {3811} (\bibinfo {year} {1998})}\BibitemShut {NoStop}%
\bibitem [{\citenamefont {Di~Rosa}(2004)}]{di_rosa_laser-cooling_2004}%
  \BibitemOpen
  \bibfield  {author} {\bibinfo {author} {\bibfnamefont {M.~D.}\ \bibnamefont
  {Di~Rosa}},\ }\bibfield  {title} {\bibinfo {title} {Laser-cooling
  molecules},\ }\href {https://doi.org/10.1140/epjd/e2004-00167-2} {\bibfield
  {journal} {\bibinfo  {journal} {Eur. Phys. J. D}\ }\textbf {\bibinfo {volume}
  {31}},\ \bibinfo {pages} {395} (\bibinfo {year} {2004})}\BibitemShut
  {NoStop}%
\bibitem [{\citenamefont {McNally}\ \emph {et~al.}(2020)\citenamefont
  {McNally}, \citenamefont {Kozyryev}, \citenamefont {Vazquez-Carson},
  \citenamefont {Wenz}, \citenamefont {Wang},\ and\ \citenamefont
  {Zelevinsky}}]{mcnally_optical_2020}%
  \BibitemOpen
  \bibfield  {author} {\bibinfo {author} {\bibfnamefont {R.~L.}\ \bibnamefont
  {McNally}}, \bibinfo {author} {\bibfnamefont {I.}~\bibnamefont {Kozyryev}},
  \bibinfo {author} {\bibfnamefont {S.}~\bibnamefont {Vazquez-Carson}},
  \bibinfo {author} {\bibfnamefont {K.}~\bibnamefont {Wenz}}, \bibinfo {author}
  {\bibfnamefont {T.}~\bibnamefont {Wang}},\ and\ \bibinfo {author}
  {\bibfnamefont {T.}~\bibnamefont {Zelevinsky}},\ }\bibfield  {title}
  {\bibinfo {title} {Optical cycling, radiative deflection and laser cooling of
  barium monohydride ({$^{\text{138}}$Ba$^{\text{1}}$H})},\ }\href
  {https://doi.org/10.1088/1367-2630/aba3e9} {\bibfield  {journal} {\bibinfo
  {journal} {New J. Phys.}\ }\textbf {\bibinfo {volume} {22}},\ \bibinfo
  {pages} {083047} (\bibinfo {year} {2020})}\BibitemShut {NoStop}%
\bibitem [{\citenamefont {Sun}\ \emph {et~al.}(2026)\citenamefont {Sun},
  \citenamefont {Dai}, \citenamefont {Koots}, \citenamefont {Riley},
  \citenamefont {Pérez-Ríos}, \citenamefont {Mitra},\ and\ \citenamefont
  {Zelevinsky}}]{sun2025chemistry}%
  \BibitemOpen
  \bibfield  {author} {\bibinfo {author} {\bibfnamefont {Q.}~\bibnamefont
  {Sun}}, \bibinfo {author} {\bibfnamefont {J.}~\bibnamefont {Dai}}, \bibinfo
  {author} {\bibfnamefont {R.}~\bibnamefont {Koots}}, \bibinfo {author}
  {\bibfnamefont {B.~C.}\ \bibnamefont {Riley}}, \bibinfo {author}
  {\bibfnamefont {J.}~\bibnamefont {Pérez-Ríos}}, \bibinfo {author}
  {\bibfnamefont {D.}~\bibnamefont {Mitra}},\ and\ \bibinfo {author}
  {\bibfnamefont {T.}~\bibnamefont {Zelevinsky}},\ }\bibfield  {title}
  {\bibinfo {title} {Chemistry in a cryogenic buffer gas cell},\ }\href
  {https://doi.org/10.1021/acs.jpclett.5c02840} {\bibfield  {journal} {\bibinfo
   {journal} {J. Phys. Chem. Lett.}\ }\textbf {\bibinfo {volume} {17}},\
  \bibinfo {pages} {640} (\bibinfo {year} {2026})}\BibitemShut {NoStop}%
\bibitem [{\citenamefont {Barry}\ \emph {et~al.}(2011)\citenamefont {Barry},
  \citenamefont {Shuman},\ and\ \citenamefont {DeMille}}]{barry_bright_2011}%
  \BibitemOpen
  \bibfield  {author} {\bibinfo {author} {\bibfnamefont {J.~F.}\ \bibnamefont
  {Barry}}, \bibinfo {author} {\bibfnamefont {E.~S.}\ \bibnamefont {Shuman}},\
  and\ \bibinfo {author} {\bibfnamefont {D.}~\bibnamefont {DeMille}},\
  }\bibfield  {title} {\bibinfo {title} {A bright, slow cryogenic molecular
  beam source for free radicals},\ }\href {https://doi.org/10.1039/C1CP20335E}
  {\bibfield  {journal} {\bibinfo  {journal} {Phys. Chem. Chem. Phys.}\
  }\textbf {\bibinfo {volume} {13}},\ \bibinfo {pages} {18936} (\bibinfo {year}
  {2011})}\BibitemShut {NoStop}%
\bibitem [{\citenamefont {Hutzler}\ \emph {et~al.}(2012)\citenamefont
  {Hutzler}, \citenamefont {Lu},\ and\ \citenamefont
  {Doyle}}]{hutzler_buffer_2012}%
  \BibitemOpen
  \bibfield  {author} {\bibinfo {author} {\bibfnamefont {N.~R.}\ \bibnamefont
  {Hutzler}}, \bibinfo {author} {\bibfnamefont {H.-I.}\ \bibnamefont {Lu}},\
  and\ \bibinfo {author} {\bibfnamefont {J.~M.}\ \bibnamefont {Doyle}},\
  }\bibfield  {title} {\bibinfo {title} {The buffer gas beam: {An} intense,
  cold, and slow source for atoms and molecules},\ }\href
  {https://doi.org/10.1021/cr200362u} {\bibfield  {journal} {\bibinfo
  {journal} {Chem. Rev.}\ }\textbf {\bibinfo {volume} {112}},\ \bibinfo {pages}
  {4803} (\bibinfo {year} {2012})}\BibitemShut {NoStop}%
\bibitem [{\citenamefont {Bulleid}\ \emph {et~al.}(2013)\citenamefont
  {Bulleid}, \citenamefont {Skoff}, \citenamefont {Hendricks}, \citenamefont
  {Sauer}, \citenamefont {Hinds},\ and\ \citenamefont
  {Tarbutt}}]{bulleid_characterization_2013}%
  \BibitemOpen
  \bibfield  {author} {\bibinfo {author} {\bibfnamefont {N.~E.}\ \bibnamefont
  {Bulleid}}, \bibinfo {author} {\bibfnamefont {S.~M.}\ \bibnamefont {Skoff}},
  \bibinfo {author} {\bibfnamefont {R.~J.}\ \bibnamefont {Hendricks}}, \bibinfo
  {author} {\bibfnamefont {B.~E.}\ \bibnamefont {Sauer}}, \bibinfo {author}
  {\bibfnamefont {E.~A.}\ \bibnamefont {Hinds}},\ and\ \bibinfo {author}
  {\bibfnamefont {M.~R.}\ \bibnamefont {Tarbutt}},\ }\bibfield  {title}
  {\bibinfo {title} {Characterization of a cryogenic beam source for atoms and
  molecules},\ }\href {https://doi.org/10.1039/C3CP51553B} {\bibfield
  {journal} {\bibinfo  {journal} {Phys. Chem. Chem. Phys.}\ }\textbf {\bibinfo
  {volume} {15}},\ \bibinfo {pages} {12299} (\bibinfo {year}
  {2013})}\BibitemShut {NoStop}%
\bibitem [{\citenamefont {Dai}\ \emph {et~al.}(2024)\citenamefont {Dai},
  \citenamefont {Sun}, \citenamefont {Riley}, \citenamefont {Mitra},\ and\
  \citenamefont {Zelevinsky}}]{dai2024laser}%
  \BibitemOpen
  \bibfield  {author} {\bibinfo {author} {\bibfnamefont {J.}~\bibnamefont
  {Dai}}, \bibinfo {author} {\bibfnamefont {Q.}~\bibnamefont {Sun}}, \bibinfo
  {author} {\bibfnamefont {B.~C.}\ \bibnamefont {Riley}}, \bibinfo {author}
  {\bibfnamefont {D.}~\bibnamefont {Mitra}},\ and\ \bibinfo {author}
  {\bibfnamefont {T.}~\bibnamefont {Zelevinsky}},\ }\bibfield  {title}
  {\bibinfo {title} {Laser cooling of a fermionic molecule},\ }\href
  {https://doi.org/10.1103/PhysRevResearch.6.033135} {\bibfield  {journal}
  {\bibinfo  {journal} {Phys. Rev. Res.}\ }\textbf {\bibinfo {volume} {6}},\
  \bibinfo {pages} {033135} (\bibinfo {year} {2024})}\BibitemShut {NoStop}%
\bibitem [{\citenamefont {Hemmerling}\ \emph {et~al.}(2016)\citenamefont
  {Hemmerling}, \citenamefont {Chae}, \citenamefont {Ravi}, \citenamefont
  {Anderegg}, \citenamefont {Drayna}, \citenamefont {Hutzler}, \citenamefont
  {Collopy}, \citenamefont {Ye}, \citenamefont {Ketterle},\ and\ \citenamefont
  {Doyle}}]{hemmerling_laser_2016}%
  \BibitemOpen
  \bibfield  {author} {\bibinfo {author} {\bibfnamefont {B.}~\bibnamefont
  {Hemmerling}}, \bibinfo {author} {\bibfnamefont {E.}~\bibnamefont {Chae}},
  \bibinfo {author} {\bibfnamefont {A.}~\bibnamefont {Ravi}}, \bibinfo {author}
  {\bibfnamefont {L.}~\bibnamefont {Anderegg}}, \bibinfo {author}
  {\bibfnamefont {G.~K.}\ \bibnamefont {Drayna}}, \bibinfo {author}
  {\bibfnamefont {N.~R.}\ \bibnamefont {Hutzler}}, \bibinfo {author}
  {\bibfnamefont {A.~L.}\ \bibnamefont {Collopy}}, \bibinfo {author}
  {\bibfnamefont {J.}~\bibnamefont {Ye}}, \bibinfo {author} {\bibfnamefont
  {W.}~\bibnamefont {Ketterle}},\ and\ \bibinfo {author} {\bibfnamefont
  {J.~M.}\ \bibnamefont {Doyle}},\ }\bibfield  {title} {\bibinfo {title} {Laser
  slowing of {CaF} molecules to near the capture velocity of a molecular
  {MOT}},\ }\href {https://doi.org/10.1088/0953-4075/49/17/174001} {\bibfield
  {journal} {\bibinfo  {journal} {J. Phys. B: At. Mol. Opt. Phys.}\ }\textbf
  {\bibinfo {volume} {49}},\ \bibinfo {pages} {174001} (\bibinfo {year}
  {2016})}\BibitemShut {NoStop}%
\bibitem [{Sup()}]{Supplemental_Material}%
  \BibitemOpen
  \bibfield  {title} {\bibinfo {title} {See {Supplemental Material} for details
  of the laser configuration}\ }\href@noop {} {}\BibitemShut {NoStop}%
\bibitem [{\citenamefont {Barry}\ \emph {et~al.}(2012)\citenamefont {Barry},
  \citenamefont {Shuman}, \citenamefont {Norrgard},\ and\ \citenamefont
  {DeMille}}]{barry_slowing_2012}%
  \BibitemOpen
  \bibfield  {author} {\bibinfo {author} {\bibfnamefont {J.~F.}\ \bibnamefont
  {Barry}}, \bibinfo {author} {\bibfnamefont {E.~S.}\ \bibnamefont {Shuman}},
  \bibinfo {author} {\bibfnamefont {E.~B.}\ \bibnamefont {Norrgard}},\ and\
  \bibinfo {author} {\bibfnamefont {D.}~\bibnamefont {DeMille}},\ }\bibfield
  {title} {\bibinfo {title} {Laser radiation pressure slowing of a molecular
  beam},\ }\href {https://doi.org/10.1103/PhysRevLett.108.103002} {\bibfield
  {journal} {\bibinfo  {journal} {Phys. Rev. Lett.}\ }\textbf {\bibinfo
  {volume} {108}},\ \bibinfo {pages} {103002} (\bibinfo {year}
  {2012})}\BibitemShut {NoStop}%
\bibitem [{\citenamefont {Norrgard}\ \emph {et~al.}(2016)\citenamefont
  {Norrgard}, \citenamefont {McCarron}, \citenamefont {Steinecker},
  \citenamefont {Tarbutt},\ and\ \citenamefont
  {DeMille}}]{norrgard_rfmot_2016}%
  \BibitemOpen
  \bibfield  {author} {\bibinfo {author} {\bibfnamefont {E.~B.}\ \bibnamefont
  {Norrgard}}, \bibinfo {author} {\bibfnamefont {D.~J.}\ \bibnamefont
  {McCarron}}, \bibinfo {author} {\bibfnamefont {M.~H.}\ \bibnamefont
  {Steinecker}}, \bibinfo {author} {\bibfnamefont {M.~R.}\ \bibnamefont
  {Tarbutt}},\ and\ \bibinfo {author} {\bibfnamefont {D.}~\bibnamefont
  {DeMille}},\ }\bibfield  {title} {\bibinfo {title} {Submillikelvin dipolar
  molecules in a radio-frequency magneto-optical trap},\ }\href
  {https://doi.org/10.1103/PhysRevLett.116.063004} {\bibfield  {journal}
  {\bibinfo  {journal} {Phys. Rev. Lett.}\ }\textbf {\bibinfo {volume} {116}},\
  \bibinfo {pages} {063004} (\bibinfo {year} {2016})}\BibitemShut {NoStop}%
\bibitem [{\citenamefont {Anderegg}\ \emph {et~al.}(2018)\citenamefont
  {Anderegg}, \citenamefont {Augenbraun}, \citenamefont {Bao}, \citenamefont
  {Burchesky}, \citenamefont {Cheuk}, \citenamefont {Ketterle},\ and\
  \citenamefont {Doyle}}]{anderegg_laser_2018}%
  \BibitemOpen
  \bibfield  {author} {\bibinfo {author} {\bibfnamefont {L.}~\bibnamefont
  {Anderegg}}, \bibinfo {author} {\bibfnamefont {B.~L.}\ \bibnamefont
  {Augenbraun}}, \bibinfo {author} {\bibfnamefont {Y.}~\bibnamefont {Bao}},
  \bibinfo {author} {\bibfnamefont {S.}~\bibnamefont {Burchesky}}, \bibinfo
  {author} {\bibfnamefont {L.~W.}\ \bibnamefont {Cheuk}}, \bibinfo {author}
  {\bibfnamefont {W.}~\bibnamefont {Ketterle}},\ and\ \bibinfo {author}
  {\bibfnamefont {J.~M.}\ \bibnamefont {Doyle}},\ }\bibfield  {title} {\bibinfo
  {title} {Laser cooling of optically trapped molecules},\ }\href
  {https://doi.org/10.1038/s41567-018-0191-z} {\bibfield  {journal} {\bibinfo
  {journal} {Nat. Phys.}\ }\textbf {\bibinfo {volume} {14}},\ \bibinfo {pages}
  {890} (\bibinfo {year} {2018})}\BibitemShut {NoStop}%
\bibitem [{\citenamefont {Jorapur}\ \emph {et~al.}(2024)\citenamefont
  {Jorapur}, \citenamefont {Langin}, \citenamefont {Wang}, \citenamefont
  {Zheng},\ and\ \citenamefont {DeMille}}]{jorapur2024srfbluemot}%
  \BibitemOpen
  \bibfield  {author} {\bibinfo {author} {\bibfnamefont {V.}~\bibnamefont
  {Jorapur}}, \bibinfo {author} {\bibfnamefont {T.~K.}\ \bibnamefont {Langin}},
  \bibinfo {author} {\bibfnamefont {Q.}~\bibnamefont {Wang}}, \bibinfo {author}
  {\bibfnamefont {G.}~\bibnamefont {Zheng}},\ and\ \bibinfo {author}
  {\bibfnamefont {D.}~\bibnamefont {DeMille}},\ }\bibfield  {title} {\bibinfo
  {title} {High density loading and collisional loss of laser-cooled molecules
  in an optical trap},\ }\href {https://doi.org/10.1103/PhysRevLett.132.163403}
  {\bibfield  {journal} {\bibinfo  {journal} {Phys. Rev. Lett.}\ }\textbf
  {\bibinfo {volume} {132}},\ \bibinfo {pages} {163403} (\bibinfo {year}
  {2024})}\BibitemShut {NoStop}%
\bibitem [{\citenamefont {Li}\ \emph {et~al.}(2024)\citenamefont {Li},
  \citenamefont {Holland}, \citenamefont {Lu},\ and\ \citenamefont
  {Cheuk}}]{li2024cafbluemot}%
  \BibitemOpen
  \bibfield  {author} {\bibinfo {author} {\bibfnamefont {S.~J.}\ \bibnamefont
  {Li}}, \bibinfo {author} {\bibfnamefont {C.~M.}\ \bibnamefont {Holland}},
  \bibinfo {author} {\bibfnamefont {Y.}~\bibnamefont {Lu}},\ and\ \bibinfo
  {author} {\bibfnamefont {L.~W.}\ \bibnamefont {Cheuk}},\ }\bibfield  {title}
  {\bibinfo {title} {Blue-detuned magneto-optical trap of {CaF} molecules},\
  }\href {https://doi.org/10.1103/PhysRevLett.132.233402} {\bibfield  {journal}
  {\bibinfo  {journal} {Phys. Rev. Lett.}\ }\textbf {\bibinfo {volume} {132}},\
  \bibinfo {pages} {233402} (\bibinfo {year} {2024})}\BibitemShut {NoStop}%
\bibitem [{\citenamefont {Hallas}\ \emph {et~al.}(2026)\citenamefont {Hallas},
  \citenamefont {Li}, \citenamefont {Vilas}, \citenamefont {Robichaud},
  \citenamefont {Anderegg},\ and\ \citenamefont {Doyle}}]{w9qc-rczf}%
  \BibitemOpen
  \bibfield  {author} {\bibinfo {author} {\bibfnamefont {C.}~\bibnamefont
  {Hallas}}, \bibinfo {author} {\bibfnamefont {G.~K.}\ \bibnamefont {Li}},
  \bibinfo {author} {\bibfnamefont {N.~B.}\ \bibnamefont {Vilas}}, \bibinfo
  {author} {\bibfnamefont {P.}~\bibnamefont {Robichaud}}, \bibinfo {author}
  {\bibfnamefont {L.}~\bibnamefont {Anderegg}},\ and\ \bibinfo {author}
  {\bibfnamefont {J.~M.}\ \bibnamefont {Doyle}},\ }\bibfield  {title} {\bibinfo
  {title} {High compression blue-detuned magneto-optical trap of polyatomic
  molecules},\ }\href {https://doi.org/10.1103/w9qc-rczf} {\bibfield  {journal}
  {\bibinfo  {journal} {Phys. Rev. Lett.}\ }\textbf {\bibinfo {volume} {136}},\
  \bibinfo {pages} {133402} (\bibinfo {year} {2026})}\BibitemShut {NoStop}%
\bibitem [{\citenamefont {Yu}\ \emph {et~al.}(2026)\citenamefont {Yu},
  \citenamefont {You}, \citenamefont {Bao}, \citenamefont {Anderegg},
  \citenamefont {Hallas}, \citenamefont {Li}, \citenamefont {Lim},
  \citenamefont {Chae}, \citenamefont {Ketterle}, \citenamefont {Ni},\ and\
  \citenamefont {Doyle}}]{yu_conveyor-belt_2026}%
  \BibitemOpen
  \bibfield  {author} {\bibinfo {author} {\bibfnamefont {S.~S.}\ \bibnamefont
  {Yu}}, \bibinfo {author} {\bibfnamefont {J.}~\bibnamefont {You}}, \bibinfo
  {author} {\bibfnamefont {Y.}~\bibnamefont {Bao}}, \bibinfo {author}
  {\bibfnamefont {L.}~\bibnamefont {Anderegg}}, \bibinfo {author}
  {\bibfnamefont {C.}~\bibnamefont {Hallas}}, \bibinfo {author} {\bibfnamefont
  {G.~K.}\ \bibnamefont {Li}}, \bibinfo {author} {\bibfnamefont
  {D.}~\bibnamefont {Lim}}, \bibinfo {author} {\bibfnamefont {E.}~\bibnamefont
  {Chae}}, \bibinfo {author} {\bibfnamefont {W.}~\bibnamefont {Ketterle}},
  \bibinfo {author} {\bibfnamefont {K.-K.}\ \bibnamefont {Ni}},\ and\ \bibinfo
  {author} {\bibfnamefont {J.~M.}\ \bibnamefont {Doyle}},\ }\bibfield  {title}
  {\bibinfo {title} {A conveyor-belt magneto-optical trap of {CaF}},\ }\href
  {https://doi.org/10.1038/s41467-025-67944-6} {\bibfield  {journal} {\bibinfo
  {journal} {Nat. Commun.}\ }\textbf {\bibinfo {volume} {17}},\ \bibinfo
  {pages} {1175} (\bibinfo {year} {2026})}\BibitemShut {NoStop}%
\bibitem [{\citenamefont {Zeng}\ \emph {et~al.}(2026)\citenamefont {Zeng},
  \citenamefont {Yang}, \citenamefont {Deng},\ and\ \citenamefont
  {Yan}}]{pd77-s994}%
  \BibitemOpen
  \bibfield  {author} {\bibinfo {author} {\bibfnamefont {Z.}~\bibnamefont
  {Zeng}}, \bibinfo {author} {\bibfnamefont {S.}~\bibnamefont {Yang}}, \bibinfo
  {author} {\bibfnamefont {S.}~\bibnamefont {Deng}},\ and\ \bibinfo {author}
  {\bibfnamefont {B.}~\bibnamefont {Yan}},\ }\bibfield  {title} {\bibinfo
  {title} {Direct loading of {BaF} molecules with a conveyor-belt
  magneto-optical trap},\ }\href {https://doi.org/10.1103/pd77-s994} {\bibfield
   {journal} {\bibinfo  {journal} {Phys. Rev. Lett.}\ }\textbf {\bibinfo
  {volume} {136}},\ \bibinfo {pages} {073402} (\bibinfo {year}
  {2026})}\BibitemShut {NoStop}%
\bibitem [{\citenamefont {Hallas}\ \emph {et~al.}(2023)\citenamefont {Hallas},
  \citenamefont {Vilas}, \citenamefont {Anderegg}, \citenamefont {Robichaud},
  \citenamefont {Winnicki}, \citenamefont {Zhang}, \citenamefont {Cheng},\ and\
  \citenamefont {Doyle}}]{hallas2023caohodt}%
  \BibitemOpen
  \bibfield  {author} {\bibinfo {author} {\bibfnamefont {C.}~\bibnamefont
  {Hallas}}, \bibinfo {author} {\bibfnamefont {N.~B.}\ \bibnamefont {Vilas}},
  \bibinfo {author} {\bibfnamefont {L.}~\bibnamefont {Anderegg}}, \bibinfo
  {author} {\bibfnamefont {P.}~\bibnamefont {Robichaud}}, \bibinfo {author}
  {\bibfnamefont {A.}~\bibnamefont {Winnicki}}, \bibinfo {author}
  {\bibfnamefont {C.}~\bibnamefont {Zhang}}, \bibinfo {author} {\bibfnamefont
  {L.}~\bibnamefont {Cheng}},\ and\ \bibinfo {author} {\bibfnamefont {J.~M.}\
  \bibnamefont {Doyle}},\ }\bibfield  {title} {\bibinfo {title} {Optical
  trapping of a polyatomic molecule in an $\ensuremath{\ell}$-type parity
  doublet state},\ }\href {https://doi.org/10.1103/PhysRevLett.130.153202}
  {\bibfield  {journal} {\bibinfo  {journal} {Phys. Rev. Lett.}\ }\textbf
  {\bibinfo {volume} {130}},\ \bibinfo {pages} {153202} (\bibinfo {year}
  {2023})}\BibitemShut {NoStop}%
\bibitem [{\citenamefont {Sawaoka}\ \emph {et~al.}(2026)\citenamefont
  {Sawaoka}, \citenamefont {Nasir}, \citenamefont {Lunstad}, \citenamefont
  {Li}, \citenamefont {Mango}, \citenamefont {Lasner},\ and\ \citenamefont
  {Doyle}}]{2h36-myzp}%
  \BibitemOpen
  \bibfield  {author} {\bibinfo {author} {\bibfnamefont {H.}~\bibnamefont
  {Sawaoka}}, \bibinfo {author} {\bibfnamefont {A.}~\bibnamefont {Nasir}},
  \bibinfo {author} {\bibfnamefont {A.}~\bibnamefont {Lunstad}}, \bibinfo
  {author} {\bibfnamefont {M.}~\bibnamefont {Li}}, \bibinfo {author}
  {\bibfnamefont {J.}~\bibnamefont {Mango}}, \bibinfo {author} {\bibfnamefont
  {Z.~D.}\ \bibnamefont {Lasner}},\ and\ \bibinfo {author} {\bibfnamefont
  {J.~M.}\ \bibnamefont {Doyle}},\ }\bibfield  {title} {\bibinfo {title}
  {{Optical trapping of SrOH molecules for dark matter and T-violation
  searches}},\ }\href {https://doi.org/10.1103/2h36-myzp} {\bibfield  {journal}
  {\bibinfo  {journal} {Phys. Rev. Res.}\ }\textbf {\bibinfo {volume} {8}},\
  \bibinfo {pages} {013100} (\bibinfo {year} {2026})}\BibitemShut {NoStop}%
\bibitem [{\citenamefont {Potvliege}\ \emph {et~al.}(2023)\citenamefont
  {Potvliege}, \citenamefont {Nicolson}, \citenamefont {Jones},\ and\
  \citenamefont {Spannowsky}}]{potvliege2023deuterium}%
  \BibitemOpen
  \bibfield  {author} {\bibinfo {author} {\bibfnamefont {R.~M.}\ \bibnamefont
  {Potvliege}}, \bibinfo {author} {\bibfnamefont {A.}~\bibnamefont {Nicolson}},
  \bibinfo {author} {\bibfnamefont {M.~P.~A.}\ \bibnamefont {Jones}},\ and\
  \bibinfo {author} {\bibfnamefont {M.}~\bibnamefont {Spannowsky}},\ }\bibfield
   {title} {\bibinfo {title} {Deuterium spectroscopy for enhanced bounds on
  physics beyond the standard model},\ }\href
  {https://doi.org/10.1103/PhysRevA.108.052825} {\bibfield  {journal} {\bibinfo
   {journal} {Phys. Rev. A}\ }\textbf {\bibinfo {volume} {108}},\ \bibinfo
  {pages} {052825} (\bibinfo {year} {2023})}\BibitemShut {NoStop}%
\end{thebibliography}


\end{document}

\title{Supplemental Material for ``Magneto-Optical Trapping of a Metal Hydride Molecule''}

\author{Jinyu Dai}
\email{jd3706@columbia.edu}
\affiliation{Department of Physics, Columbia University, New York, New York 10027, USA}

\author{Benjamin Riley}
\affiliation{Department of Physics, Columbia University, New York, New York 10027, USA}

\author{Qi Sun}
\affiliation{Department of Physics, Columbia University, New York, New York 10027, USA}

\author{Debayan Mitra}
\affiliation{Department of Physics, Columbia University, New York, New York 10027, USA}
\affiliation{Department of Physics, Indiana University, Bloomington, Indiana 47405, USA}

\author{Tanya Zelevinsky}
\affiliation{Department of Physics, Columbia University, New York, New York 10027, USA}

\date{\today}

\maketitle

\section{S1. Laser configuration}

\begin{table*}[h!]
\begin{threeparttable}
\centering
\setlength{\tabcolsep}{4pt}
\caption{Laser frequencies used in this work. The values for the slowing lasers represent their center frequencies ($^{\ast}$), and the $^{+/-}$ notation is used to distinguish the ($F=1$) states for different $J$ states. For velocity sensitive detection, the resonant laser frequencies are displayed. The shown MOT laser frequencies are detuned by $-5$~MHz from resonance, the optimal MOT detuning.}
\begin{tabular}{c c c c c c c c c c c c}
\hline\hline
 & Ground & $\nu$ & $N$ & $J$ & $F$ & Excited & $\nu'$ & $N'$ & $J'$ & $F'$ & Frequency (THz) \\
\hline
\multirow{3}{*}{Slowing}
 & $X$ & $0$ & $1$ & $1/2$, $3/2$ & $0$, $1^+$, $1^-$, $2$ & $A$ & $0$ & -- & $1/2$ & $0$, $1$ & $431.275351^{\ast}$ \\
 & $X$ & $1$ & $1$ & $1/2$, $3/2$ & $0$, $1^+$, $1^-$, $2$ & $B$ & $0$ & $0$  & $1/2$ & $0$, $1$ & $434.255716^{\ast}$ \\
 & $X$ & $2$ & $1$ & $1/2$, $3/2$ & $0$, $1^+$, $1^-$, $2$ & $A$ & $1$ & -- & $1/2$ & $0$, $1$ & $395.717941^{\ast}$ \\
\hline
\multirow{7}{*}{\parbox[c]{2cm}{Velocity\\sensitive\\detection}}  % New section with 7 rows
 & $X$ & $0$ & $1$ & $1/2$ & $0$ & $A$ & $0$ & -- & $3/2$ & $1$, $2$ & $431.604611$ \\
 & $X$ & $0$ & $1$ & $1/2$ & $1$ & $A$ & $0$ & -- & $3/2$ & $1$, $2$ & $431.604664$ \\
 & $X$ & $0$ & $1$ & $3/2$ & $1$ & $A$ & $0$ & -- & $3/2$ & $1$, $2$ & $431.602751$ \\
 & $X$ & $0$ & $1$ & $3/2$ & $2$ & $A$ & $0$ & -- & $3/2$ & $1$, $2$ & $431.602651$ \\
 & $X$ & $0$ & $3$ & $5/2$ & $2$ & $A$ & $0$ & -- & $3/2$ & $1$, $2$ & $430.338953$ \\
 & $X$ & $0$ & $3$ & $5/2$ & $3$ & $A$ & $0$ & -- & $3/2$ & $1$, $2$ & $430.339023$ \\
 & $A$ & $0$ & -- & $3/2$ & $1$, $2$ & $E$ & $0$ & -- & $3/2$ & $1$, $2$ & $179.685381$ \\
\hline
\multirow{4}{*}{MOT}
 & $X$ & $0$ & $1$ & $1/2$ & $0$ & $A$ & $0$ & -- & $1/2$ & $1$ & $431.276545$ \\
 & $X$ & $0$ & $1$ & $1/2$ & $1$     & $A$ & $0$ & -- & $1/2$ & $1$ & $431.276598$ \\
 & $X$ & $0$ & $1$ & $3/2$ & $1$ & $A$ & $0$ & -- & $1/2$ & $0$ & $431.274667$ \\
 & $X$ & $0$ & $1$ & $3/2$ & $2$     & $A$ & $0$ & -- & $1/2$ & $1$ & $431.274585$ \\
\hline\hline
\end{tabular}%
\label{tab:lasers}
\end{threeparttable}
\end{table*}
Laser frequencies used in this work are summarized in Table~\ref{tab:lasers}. The slowing lasers are frequency broadened and the given values represent their center frequencies. For the main cycling laser of slowing [$A^2\Pi_{1/2}(\nu'=0,\ J'=1/2,\ +)\leftarrow X^2\Sigma^{+}(\nu=0,\ N=1,\ -)$] we use electro-optic modulators (EOMs) operating at $947.1$~MHz and $50.83$~MHz to cover the spin-rotation and hyperfine splittings in the ground state, using the $\pm1$ orders of the sidebands. The hyperfine EOM, combined with the white-light slowing EOM operating at $4.335$~MHz, generates a frequency broadening of $402$~MHz. The two repumping lasers share another set of EOMs operating at $942$~MHz, $53.5$~MHz, and $4.185$~MHz, corresponding to the spin-rotation splitting, hyperfine splitting, and white-light generation. Frequency broadening for the repumping lasers is $424$~MHz. Table~\ref{tab:lasers} also shows the resonant laser frequencies for velocity sensitive detection. For our two-photon background-free detection scheme, the velocity information is obtained via detuning the $E\leftarrow A$ laser from resonance. The shown MOT laser frequencies are detuned by $-5$~MHz from resonance. All frequencies are measured with a HighFinesse WS7-60 wavemeter, which has a systematic uncertainty of $\sim$$60$~MHz.

%%%%%%%%%%%%%%%%%%%%%%%%%%%%%%% References %%%%%%%%%%%%%%%%%%%%%%%%%%%%%%%

% \bibliographystyle{apsrev4-2}
% \bibliography{references}
